\documentclass[manuscript=article]{achemso}

\usepackage[version=3]{mhchem} 
\usepackage[T1]{fontenc}       
\usepackage{graphicx}
\usepackage{amsmath}
\usepackage{amssymb}
\usepackage{bm}
\usepackage{textcomp}
\usepackage[english]{babel}
\usepackage[utf8]{inputenc}
\usepackage{times}
\usepackage{cancel}
\usepackage{color}
\usepackage{here}
\usepackage{amsfonts}
\usepackage[percent]{overpic}
\usepackage{hyperref}
\usepackage{url}
\usepackage{mathpazo}

\author{Andr\'{e}s~C\'{o}rdoba}
\email{andcorduri@gmail.com}
\affiliation[Universidad de Concepción]
{Department of Chemical Engineering, Universidad de Concepción, Concepción, Chile}

\title[Interplay between motor and Brownian forces]
  {The Effects of the Interplay Between Motor and Brownian Forces on the
Rheology of Active Gels}

\abbreviations{IR,NMR,UV}

\begin{document}



%
%
%


\begin{abstract}
Active gels perform key mechanical roles inside the cell, 
such as cell division, motion and force sensing. 
The unique mechanical properties required to perform such functions arise 
from the interactions between molecular motors and 
semi-flexible polymeric filaments. Molecular motors can convert the energy
released in the hydrolysis of ATP into forces of up to pico-Newton magnitudes.  
Moreover, the polymeric filaments  that form active gels
are flexible enough to respond to Brownian forces, but also stiff enough
to support the large tensions induced by the motor-generated forces. \textcolor{black}{
Brownian forces are expected to have a significant effect especially at motor activities at 
which stable non-contractile {\it in vitro} active gels are prepared for rheological measurements.}
Here, a microscopic mean-field theory of active gels  
originally formulated in the limit of motor-dominated dynamics
is extended to include Brownian forces. 
\textcolor{black}{In the model presented here Brownian forces are included
accurately, at real room temperature, even in systems with high motor activity.}
It is shown that a subtle interplay, or competition, between
motor-generated forces and Brownian forces has an important
impact in the mass transport and rheological properties of active gels. 
The model predictions show that at low frequencies the dynamic modulus of active gels
is determined mostly by motor protein dynamics. However, Brownian forces
significantly increase the breadth of the relaxation spectrum and can
affect the shape of the dynamic modulus over a wide frequency range
even for ratios of motor to Brownian forces of more than a hundred.
Since the ratio between motor and Brownian forces 
is sensitive to ATP concentration, the results presented here shed some light on how the 
transient mechanical response of active gels changes with varying ATP concentration.
\end{abstract}

\noindent This article may be downloaded for personal use only. 
Any other use requires prior permission of the author and ACS Publications. 
This article appeared in A. C\'{o}rdoba, 
J. Phys. Chem. B 2018, 122, 15, 4267--4277
and may be found at: \\ \url{https://doi.org/10.1021/acs.jpcb.8b00238}.

\section{Introduction}\label{INTRO}

Active gels are networks of semiflexible polymer filaments driven by motor proteins
that can convert chemical energy from the hydrolysis of adenosine triphosphate (ATP)
to mechanical work and motion. Active gels perform essential functions in 
living tissue including motions, generation of forces and sensing of external forces.
Moreover,  active gels play a central role in driving cell division and 
cell motility \citep{e2011active, surrey2001physical, murrell2012f}. 
For instance, a widely studied active gel is the actin cortex which is a
disordered network of F-actin decorated with myosin II motors. 
Changes in cell shape, as required for migration and division, are mediated by the cell cortex.
Myosin II motors drive contractility of the cortical actin network, enabling shape change and 
cytoplasmic flows underlying important physiological processes such as cell
division, migration and tissue morphogenesis \citep{murrell2012f}. Moreover,
Actin filaments are enriched beneath the plasma membrane,
especially at the pre- and postsynaptic regions in neurons.
Among the myosin superfamily proteins, myosin
Va, myosin Vb, and myosin VI are primarily involved in transport
in the synaptic regions \citep{hirokawa2010molecular}. Myosin II is also involved
in dynamic organization of actin bundles in the postsynaptic
spines and is related to synaptic plasticity through control of
spine shape \citep{hirokawa2010molecular}.

The semiflexible filaments that form active gels, such as 
actin and tubulin, are characterized 
by having a persistence length (length over which the 
tangent vectors to the contour of the filament
remain correlated) that is larger than the mesh size of the network, 
but typically smaller than the contour length of the filament. For instance,
for filamentous actin (F-actin)
the persistence length is around $10~\mu$m, while the mesh size of 
actin networks is estimated to be approximately 
$1~\mu$m. \cite{marko1995stretching, gittes1998dynamic, storm2005nonlinear}
In active gels, molecular motors assemble into cylindrical aggregates
with groups of binding heads at each end that can attach 
to active sites along the semiflexible filaments that constitute the gel 
\citep{huxley1974muscular, hill1975some}. For example, in actomyosin gels 
myosin II motors ensemble into cylindrical aggregates 
of about $1.5~\mu$m in length, each aggregate has around $400$ 
attachment heads \cite{murrell2012f}. 
The binding heads of motor proteins can attach to the polymeric filaments that form
the gel, in the absence of ATP motor clusters act as passive cross-links 
between the polymeric filaments that constitute the gel. Moreover, these 
heads in the motor proteins can also bind to ATP. When ATP molecules are present the 
heads of the motor proteins to which ATP binds will detach from the site on the filament 
to where they where attached. Using the chemical energy from the hydrolysis 
of ATP the detached motor heads will move towards the next attachment site 
along the filament contour and reattach. The direction in which motors move is determined 
by the filament structural polarity \citep{yamada1993movement, julicher1995cooperative}. 
In active gels, motor clusters have several binding heads performing this same process. 
The binding heads on one end of the motor cluster can
be walking along a particular filament while the binding heads on the other end
of the motor cluster can be attached to a different, neighboring, filament. 
This second filament will feel a force due to the motion
of the motor along the first filament.  For instance,
in actomyosin gels approximately four out of 400 attachment heads 
in a myosin motor cluster are attached to an actin filament at any given time.  
Each motor head can exert $3-4$ pN force, assuming that each motor head 
contributes force additively, then the whole myosin aggregate can exert roughly 
$12-16$ pN of force \cite{murrell2012f} on an actin filament. 

Active gels exhibit modulated flowing phases and a macroscopic 
phase separation at high activity \cite{voituriez2006generic, ndlec1997self}. 
However at low ATP or motor protein concentrations stable active gels 
have been successfully prepared {\it in vitro} to study their mechanical and 
rheological properties \cite{stuhrmann2012nonequilibrium,
bertrand2012active, mizuno2007nonequilibrium, sanchez2012spontaneous}.
The rheological properties of active gels have also been studied {\it in vivo}.
For instance the motility of {\it Amoeba proteus} was examined using the technique of passive 
particle tracking microrheology \cite{rogers2008intracellular}. 
Endogenous particles in the amoebae cytoplasm were tracked,
which displayed subdiffusive motion at short timescales, corresponding to Brownian motion
in a viscoelastic medium, and superdiffusive motion at long timescales due
to the convection of the cytoplasm. The subdiffusive motion of tracer particles
was characterized by a rheological scaling exponent of $3/4$ in the cortex, 
indicative of the semiflexible dynamics of the actin fiber.
The mechanical properties of  embryos of {\it Drosophila} in early stages of development 
have also been recently studied using microrheology \cite{wessel2015mechanical}.

These rheological experiments in active polymeric networks
have revealed fundamental differences from their passive counterparts.
For instance, it has been observed that the
fluctuation-dissipation theorem (FDT) is violated in active gels \citep{bertrand2012active, 
stuhrmann2012nonequilibrium, mizuno2007nonequilibrium}. This violation 
shows up as a frequency-dependent  discrepancy between the material response function
obtained from driven and passive microrheology experiments. 
The discrepancy is largest at low frequencies, around 1 Hz, but it disappears at 
larger frequencies, around 100 Hz. In driven microrheology an external force is applied to 
the probe bead and the material response function is calculated 
from the bead position signal. On the other hand, in passive microrheology experiments,
no external force is applied, and the material response function is calculated from the
bead position autocorrelation function using the FDT \citep{mizuno2007nonequilibrium}.

Recently a model of active gels that can
predict this violation of the FDT in active gels was proposed \cite{Cordoba2014, cordoba2015role}.
In the active single-chain model, molecular motors are accounted for by using a mean-field 
approach and the stochastic state variables evolve according  to a proposed differential 
Chapman-Kolmogorov equation. Initially the model was introduced with a level 
of description that had the minimum set of variables necessary to describe 
some of the most characteristic mechanical and rheological properties of active gels 
\citep{Cordoba2014}. For instance, only the simplified 
case of dumbbells was considered, meaning that only two motor attachment sites per filament 
were allowed.  The filaments were modeled as Fraenkel springs, and the motor force 
distribution was made a Dirac delta function centered around a 
mean motor stall force. With those assumptions it was possible to obtain
closed-form analytical expressions for several observables of the model, 
such as relaxation modulus and fraction of buckled filaments. 
Later, some of the assumptions made in the original model were relaxed 
\citep{cordoba2015role}. More specifically, bead-spring chains with multiple
attachment sites, finite-extensibility of the filament segments and
experimentally-observed motor force distributions were considered. 
In that form the model can not longer be
solved analytically and numerical simulations are employed.

An important simplification in previous versions of the active 
single-chain model was that Brownian forces were neglected. 
This simplification was assumed to be appropriate when
motor-generated forces are much larger than Brownian forces.
However, the model without Brownian forces 
rendered only a qualitative description at ATP concentrations at which stable 
non-contractile {\it in vitro} active gels are prepared for rheological measurements.
The main reason for excluding Brownian forces from the original single-chain 
active model was to draw a clear distinction between the proposed model 
and a widespread approach for the mesoscopic modeling of active 
systems which utilizes a so-called effective temperature. 
In that approach the motor generated forces  are modeled as 
Brownian forces, but an effective temperature is introduced, 
which is higher than the real temperature 
and is meant to account for the larger magnitude of the non-equilibrium fluctuations
\cite{liverpool2001viscoelasticity, PhysRevLett.97.268101, liverpool2007bridging, 
palacci2010sedimentation}. In this work Brownian forces 
are included in the active single-chain model, but is
important to emphasize that no effective temperature is introduced  
and the dynamics of the motor-generated forces are treated in
the same way that was done in previous versions of the model.
\textcolor{black}{This allows for the Brownian forces to be modeled 
accurately, at real room temperature, even in systems with high motor activity.}

The purpose of this work is to provide molecular-level
insight into how the interplay, or competition, between Brownian 
forces and motor-generated forces influence the mass transport and 
rheological properties of active gels. For this purpose the 
single-chain model for active gels is extended to include 
Brownian forces. The paper is organized as follows.
It begins with a brief description of the active single-chain model
with Brownian forces, discussing the main assumptions, and parameters. 
An outline of the numerical procedure used to solve the model is also
provided there. This is followed by a subsection where the 
predictions of the active single-chain model for the mass transport mechanisms
of filaments in active gels are presented and discussed. 
It is shown that the model predictions have good qualitative agreement
with observations in microtubule and kinesin solutions.
This is followed by a subsection where the dependence 
of the dynamic modulus of active gels on the
ratio between motor and Brownian forces is discussed. 
Also the creep compliance predicted by the active single-chain model with 
Brownian forces is compared to published microrheology data for actomyosin gels.

\section{Results and discussion}

\subsection{The active single-chain model}\label{MODEL}

The active single-chain model follows a probe filament
and approximate its surroundings by an 
effective medium of motor clusters that attach and detach from 
specific sites along the probe filament. The motors are assumed to
form pair-wise interactions between filaments. When a motor is attached to the
probe filament it is stepping forward on
another filament in the mean-field and therefore pulls/pushes on the probe filament. 
To model these interactions a mean-field approach is employed, in which filaments have
prescribed probabilities to undergo a transition from one attachment/detachment state into 
another depending on the state of the particular filament.
\textcolor{black}{A diagram of the active single-chain model is shown in Fig. 1}.

\begin{figure}[h t]
\vspace{5mm}
\includegraphics[width=0.85\linewidth]{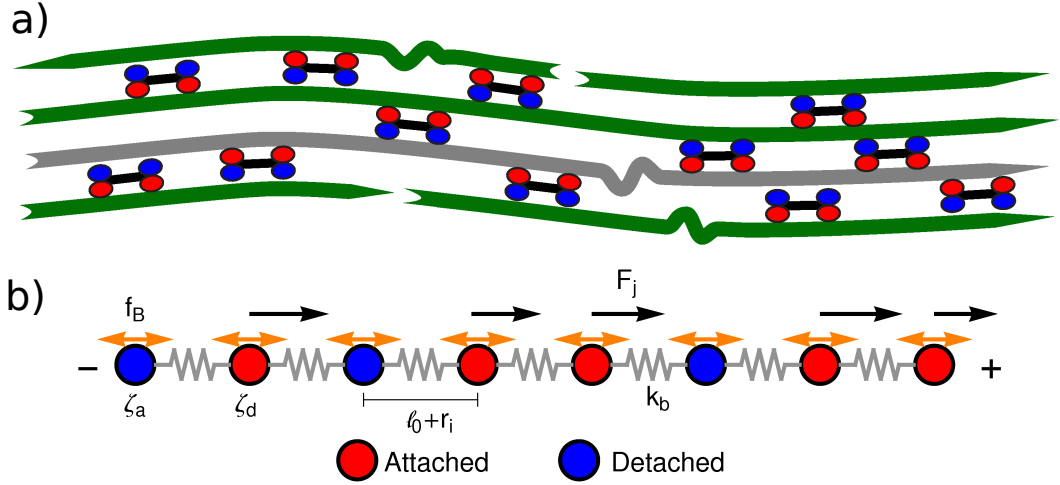}
\caption{Sketch of the active single-chain model with Brownian forces. 
a) Active bundle formed by polar filaments and motors (which can move towards the barbed
end of the filaments). Motors attach and detach from the filaments. After detaching
from a given filament a motor will step forward in that filament and will exert a force on
the other filament where it is still attached. The gray filament indicates a
probe filament whose dynamics are followed by the model. b) The probe
filament is represented by a bead-spring chain.
Red beads represent attachment sites in the filament where a motor is attached, 
$\zeta_a$ is the friction coefficient of those beads.
Blue beads represent sites in the filaments where no motor is attached
$\zeta_d$ is the friction coefficient of those beads. The orange double-headed
arrows represent the Brownian forces, $f_{\rm B}$, acting on the beads.
$F_j$ is a motor-generated force acting on bead $j$. Motors
generate a force on the filament only when attached. 
$\ell_0$ is the rest length of the strands before addition of ATP; $r_i$ is the
change in the end-to-end distance of a strand due to Brownian 
and motor-generated forces.}
\label{fig0}
\end{figure}

The attachment state is represented by a single number $s$; 
by allocating number $0$ to free sites and $1$ to sites attached to a motor.
In the following, the number $0$ or $1$ assigned to site $j$ on a filament in an 
attachment state $s$ is denoted by $n_j(s)$. By definition, $s$ takes one of the values
$0, 1, ..., 2^N(= s_{\rm max})$, where $N$ is the total number of sites
in the probe filament. The network strands have contour length $\ell_c$, 
therefore a filament has total contour length length $\ell_f=(N-1)\ell_c$.
Before addition of ATP all the motor clusters are attached (as passive cross-links) 
and the strands between them have relaxed end-to-end length $\ell_0$. 
\textcolor{black}{This is the rest length of the filament segments and therefore there is no  
tension in the filaments before addition of ATP. 
For instance, if an actin network is formed in the presence 
of high concentrations of myosin
the resulting cross-link density is higher and $\ell_0$
smaller than in the same network formed under lower concentrations.
In typical actin networks prepared {\it in vitro} $\ell_0$
is on the order of $1~\mu$m \citep{gittes1998dynamic, storm2005nonlinear}}.
The average time a motor spends attached to a site before detaching from
it is labeled $\tau_d$, whereas the average time a motor spends detached before 
reattaching is given by the model parameter $\tau_a$. 
The force generated by a motor attached to site $j$ will be denoted $F_j$. 
Molecular motors can only move in one direction along the filament,
determined by the filament's polarity. Filaments are thus 
expected to move in the opposite direction. In this single-chain description 
this is described by making all the forces $F_j$, that the motors exert 
on the sites of the probe filament, have the same sign (either positive or negative).

Another force acting on the filament is the viscous 
drag from the surrounding solvent. The frictional force from the surrounding solvent 
is characterized by a friction coefficient.
Motor clusters (i.e.: $1.5\mu$m for myosin II thick filaments) 
increase the friction coefficient of the filament when attached to an active site.
Therefore this friction coefficient is allowed to
take two different values: $\zeta_a$ when attached,
and $\zeta_d<\zeta_a$ if there is no motor attached to that site.  

In summary the following state variables are used to construct the 
model of the active gel $\Omega:\{s,\bm{F}, \bm{r}\}$.
Where $\bm{F}:=\{F_1, F_2, F_j,... F_N\}$ is a vector that contains the 
motor forces for all the sites and $\bm{r}:=\{r_1, r_2, r_i,... r_{N-1}\}$ 
is a vector that contains the change from the rest length in the 
end-to end distance of all the strands. Now let $\psi(\Omega)$ be the distribution 
function describing the probability of finding an active filament in state $s$ with strands 
with a change in their end-to-end distance 
$\bm{r}$ due to motor forces $\bm{F}$ at time $t$. 
The time evolution for $\psi(\Omega)$ is given by the following differential
Chapman-Kolmogorov equation:
\begin{eqnarray}\label{CKeq}
\frac{\partial \psi(\hat{\omega},\bm{r};t)}{\partial t}&=&\sum_{i,j=1}^{N-1}
\frac{\partial}{\partial r_i}\bigg\{\psi(\hat{\omega},\bm{r};t)\bigg[-\dot{\epsilon}(t) r_j+A_{i,j}(s)f(r_j)-
\frac{\left(F_{j+1}-F_j\right)\delta_{i,j}}{\zeta_a}\bigg]\\
\nonumber
&+&A_{i,j}(s)k_{\rm B}T\dfrac{\partial\psi(\hat{\omega},\bm{r};t)}{\partial r_j}\bigg\}
+\sum_{s'=0}^{s_{\rm max}}\int\mathbb{W}_{s,s'}(\hat{\omega}|\hat{\omega}')
\psi(\hat{\omega}',\bm{r};t)d\bm{F}'
\end{eqnarray}
where $\hat{\omega}:\{s,\bm{F}\}$ is a subspace of $\Omega$,
$\epsilon(t)$ is an externally applied strain and $f(r_j)$ is the spring force. The
term proportional to $k_{\rm B}T$, which introduces the
effect of Brownian forces, differentiates this model from previous versions 
of the active single-chain model which did not include it \citep{Cordoba2014, cordoba2015role}.

For the Fraenkel springs considered here $f(r_j)=-k_b r_j$. The linear spring constant for 
inextensible (ie.: fixed contour length $\ell_c$) semiflexible filaments
\citep{mackintosh1995elasticity} is given by $k_b=\dfrac{90 k_{\rm B} T \ell^2_p}{\ell_0^4}$.
The rest length $\ell_0$ is related to the contour length by $\ell_0=\ell_c-\dfrac{\ell_0^2}{6\ell_p}$.
For F-actin filaments, the persistence length, $\ell_p$, is approximately 10 $\mu$m 
and $\ell_0\sim1\mu$m and therefore $k_b$ is on the order of $1~\mu$N/m.
The matrix $A_{i,j}(s)$, in eq.(\ref{CKeq}) gives the connectivity of the sites as
a function of the motor-attachment state, $s$, and is defined as,
\begin{eqnarray}\label{Af}
A_{i,j}(s)=-a_i(s)\delta_{i,j-1}+[a_i(s)+b_i(s)]\delta_{i,j}-b_i(s)\delta_{i,j+1}
\end{eqnarray}
where $a_i(s)$ and $b_i(s)$ are given by,
\begin{eqnarray}\label{ab}
\left\{a_i(s), b_i(s)\right\}=
\left\{ \begin{array}{lll} 
\left\{\frac{1}{\zeta_a}, \frac{1}{\zeta_a}\right\} 
& {\rm if} & \{n_{i}(s),n_{i+1}(s)\}=\{1,1\} \\
\left\{\frac{1}{\zeta_d}, \frac{1}{\zeta_a}\right\} 
& {\rm if} & \{n_{i}(s),n_{i+1}(s)\}=\{0,1\} \\
\left\{\frac{1}{\zeta_a},\frac{1}{\zeta_d}\right\} & {\rm if} 
& \{n_{i}(s),n_{i+1}(s)\}=\{1,0\} \\
\left\{\frac{1}{\zeta_d},\frac{1}{\zeta_d}\right\} 
& {\rm if} & \{n_{i}(s),n_{i+1}(s)\}=\{0,0\}. \\
\end{array}
\right.
\end{eqnarray}
As stated above, $\zeta_a$ is the friction coefficient of a site when a motor is attached to
it and $\zeta_d$ is the friction coefficient when there is no motor attached to the site.

The transition rate matrix $\mathbb{W}(\hat{\omega}'|\hat{\omega})$ in eq.(\ref{CKeq})  
contains the transition rates between attachment/detachment states. To construct 
$\mathbb{W}(\hat{\omega}'|\hat{\omega})$ a matrix $\mathbb{K}(l)$ 
of dimensions $2^l\times2^l (l=1,...,N)$ 
is first generated by the following iterative procedure:
\begin{eqnarray}\label{TM0}
\mathbb{K}(l)=\left( \begin{array}{cc}
\mathbb{K}(l-1) & \dfrac{1}{\tau_d} \prod_{i=1}^N\delta(F_i') \bm{\delta}(2^{l-1})\\
\dfrac{p(F_{l}')}{\tau_a}\prod_{(i\neq l)=1}^N\delta(F_i')
\bm{\delta}(2^{l-1}) & \mathbb{K}(l-1)
\end{array}\right).
\end{eqnarray}
Where $\mathbb{K}(0)=1$, $\delta(...)$ is the Dirac delta function and 
$\bm{\delta}(2^{l})$ is an identity matrix of dimensions $2^l \times 2^l$. 
Then $\mathbb{W}(\hat{\omega}'|\hat{\omega})$ is defined in terms of $\mathbb{K}(l)$ as,
\begin{eqnarray}\label{TM}
\mathbb{W}(\hat{\omega}'|\hat{\omega})=\left\{\begin{array}{lll}
\mathbb{K}_{s',s}(N) & {\rm if} & s' \neq s \\
-\sum_{s''(\neq s)=0}^{s_{\rm max}}\mathbb{K}_{s'',s}(N)  & {\rm if} & s = s'.
\end{array}\right.
\end{eqnarray}
The block matrix at the upper-left or lower-right block element of 
$\mathbb{W}(\hat{\omega}'|\hat{\omega})$ represents the transition rate matrix 
of a chain having $N-1$ sites, whereas the upper right and lower left elements stand for the
detachment and attachment rates of motors in the $Nth$ site, respectively.
In the strong attachment case, when $\tau_a\ll\tau_d$, 
a good approximation is to consider only attachment states with a maximum of 
one detached motor \citep{indei2010linear}. 
\textcolor{black}{This reduces the size of the transition 
matrix from $2^N \times 2^N$ to $(N+1) \times (N+1)$. 
This approximation can make numerical simulations of the model more efficient.
Explicit forms of the transition matrices for $N=2$ and $N=3$ that illustrate this 
point are given in a Supplementary Information file. In actomyosin gels \citep{lenz2012contractile}
the typical value for the ratio $\tau_a/\tau_d$ lies around $0.005$. In this 
work the strong attachment approximation was employed in all the numerical simulations
of the model where $\tau_a/\tau_d<0.05$.} 

The function $p(F)$ is the probability
distribution from which a motor force is drawn every time a motor attaches to a site.
Motor force distributions have been measured experimentally in actomyosin
bundles \cite{thoresen2011reconstitution, lenz2012contractile}.
In order to obtain analytic solutions of the model it is 
necessary to assume that the motor force distribution
is given by a Dirac delta function centered around the mean-motor stall
force that is $p(F)=\delta(F-F_m)$. where $F_m$ is the mean motor-stall force.
A more realistic shape of these motor force distributions 
can be incorporated into the model. 
For instance, the shape of the cumulative probability 
function of myosin motor forces in actin bundles appears to
follow a gamma distribution \citep{thoresen2011reconstitution, cordoba2015role}. 
Therefore to incorporate a more realistic distribution of
motor forces in the model a fit of a gamma distribution to the experimental 
data in actomysoin bundles \citep{thoresen2011reconstitution} is used. 
\textcolor{black}{The fit to the experimental data has been shown in a previous 
paper \citep{cordoba2015role} and results} in a gamma distribution with scale parameter
$\beta=0.42 F_m$ and shape parameter of $\alpha=F_m/\beta=2.4$.

It was previously shown that closed-form 
analytic solutions to the active single-chain model can be obtain 
in certain simplified cases \citep{Cordoba2014}. 
To perform numerical simulations of the proposed model 
is more convenient to write the dynamics in phase space 
instead of the configuration space given by the Chapman-Kolmogorov 
equation. For the model with Brownian forces, the evolution equations for the 
filament segments in the phase space are given by the following stochastic 
differential equations,
\begin{eqnarray}
d r_{i,s} &=& \dot{\epsilon} r_{i,s} dt -\sum_{j=1}^{N-1}\left\{A_{i,j}(s)f(r_{j,s})-
\frac{\left[F_{j+1}-F_{j}\right]\delta_{i,j}}{\zeta_a}\right\}dt\\
\nonumber 
&+&\sqrt{2k_{\rm B}T} \sum_{j=1}^{N-1} B_{i,j}(s) dW_j(t), (i=1,2,3,...,N-1).
\end{eqnarray}
Where $dW(t)$ is a Wiener increment with statistics, 
$\left\langle {d W}_{i}(t)\right\rangle_{\rm eq}=0$ and 
$\left\langle {dW}_{i}(t){dW}_{j}(t')\right\rangle_{\rm eq}
=\delta(t-t')\delta_{ij}dtdt'$. On the other hand
the matrix $\bm{B}(s)$ must satisfy, 
$[\bm{B}(s)]\cdot[\bm{B}(s)]^{\top}=\bm{A}(s)$ which guarantees that the 
Brownian forces satisfy the fluctuation dissipation theorem. 
An efficient way to construct the matrix  $\bm{B}$ is,
\begin{eqnarray}
B_{i,j}(s)=
\left\{ \begin{array}{ccc} 
-\sqrt{2k_{\rm B} T a_i(s)}
& {\rm if} & i=j \\
\sqrt{2k_{\rm B} T b_i(s)} 
& {\rm if} & i+1=j \\
0 & {\rm elsewhere} & \\
\end{array}
\right.
\end{eqnarray}

Where $a_i(s)$ and $b_i(s)$ where defined in eq. (\ref{ab}). The evolution equation
for the state variables $s$ and $\bm{F}$ can be obtained by integrating 
eq. (\ref{CKeq}) over $\bm{r}$,
\begin{eqnarray}\label{evs}
\frac{\partial \phi(\hat{\omega};t)}{\partial t}=
\sum_{s'=0}^{s_{\rm max}}\int\mathbb{W}_{s,s'}(\hat{\omega}|\hat{\omega}')
\phi(\hat{\omega}';t)d\bm{F}'
\end{eqnarray}
where $\phi(\hat{\omega};t):=\int \psi(\hat{ \omega},\bm{r};t)d\bm{r}$.

In a numerical simulation of the model, in every time step the probability of 
jumping from a given attachment state to a a new one is calculated as 
$p=1-\exp\left\{-\Delta t\sum_{s=0}^{s_{\rm max}}\sum_{s'(\neq s)=0}^{s_{\rm max}}
\mathbb{W}_{s,s'}\right\}$. The time step of the simulation, $\Delta t$, is chosen such that
$\Delta t<1/\sum_{s=0}^{s_{\rm max}}
\sum_{s'(\neq s)=0}^{s_{\rm max}}\mathbb{W}_{s,s'}$, 
$\Delta t\ll\tau_{r,a}$ and $\Delta t\ll\tau_{r,d}$, where 
$\tau_{r,a}=\zeta_a/k_b$ and $\tau_{r,d}=\zeta_d/k_b$.
If a jump is accepted then the probabilities
of a filament that is in attachment state $s'$ to jump to
attachment state $s$ are calculated according to the transition 
matrix $\mathbb{W}_{s,s'}$. If the larger probabilities correspond to 
states that require a detachment then 
one of those states is chosen randomly, the transition occurs, 
and the respective motor forces, $F_j$, are set to zero. 
If the larger probabilities correspond to states that require
an attachment then random motor forces are generated from a gamma 
distribution. Then the probability for each of those generated forces is calculated
from the gamma probability density function and the
state with the highest probability is chosen.

\subsection{Mass transport}\label{TP}

The diffusion of tracer beads or labeled filaments
inside active gels has been studied both experimentally and
theoretically \cite{levine2009mechanics, head2010nonlocal, 
stuhrmann2012nonequilibrium}. The mass transport of microscopic probe particles 
in active gels has been observed to exhibit super-diffusive behavior at time
scales at which in passive polymeric networks probe particles exhibit diffusive 
behavior \cite{levine2009mechanics, stuhrmann2012nonequilibrium, sanchez2012spontaneous}. 
For example, the mean-squared displacement, MSD, of microscopic tracer
particles embedded in microtubules and kinesin solutions for increasing ATP 
concentrations from 0 to 5.6 mM has been reported 
\cite{sanchez2012spontaneous}. At ATP concentrations above 73 $\mu$M the the MSD exhibits
ballistic behavior at all lag times measured, from $10^{-1}$ to $10^3$ seconds. 
MSDs of endogenous particles inside the cytoplasm of live and motile {\it Amoeba proteus}
\cite{rogers2008intracellular} have also been reported. In that system, the MSD exhibits
either diffusive or super-diffusive regions at lag times above $\sim 3$ ms , 
depending on the location, inside the amoeba, of the tracked particle.
Moreover, Head et al. \cite{head2011microscopic} have also 
investigated the mass-transport of filaments in active gels using a multi-chain simulation
of an active gel. They found that filament translational motion ranges from diffusive to 
super-diffusive, depending on the ratio of attachment/detachment rates of the motors
\cite{head2011microscopic}.

\begin{figure}[h t]
\vspace{5mm}
\begin{overpic}[width=0.48\linewidth]{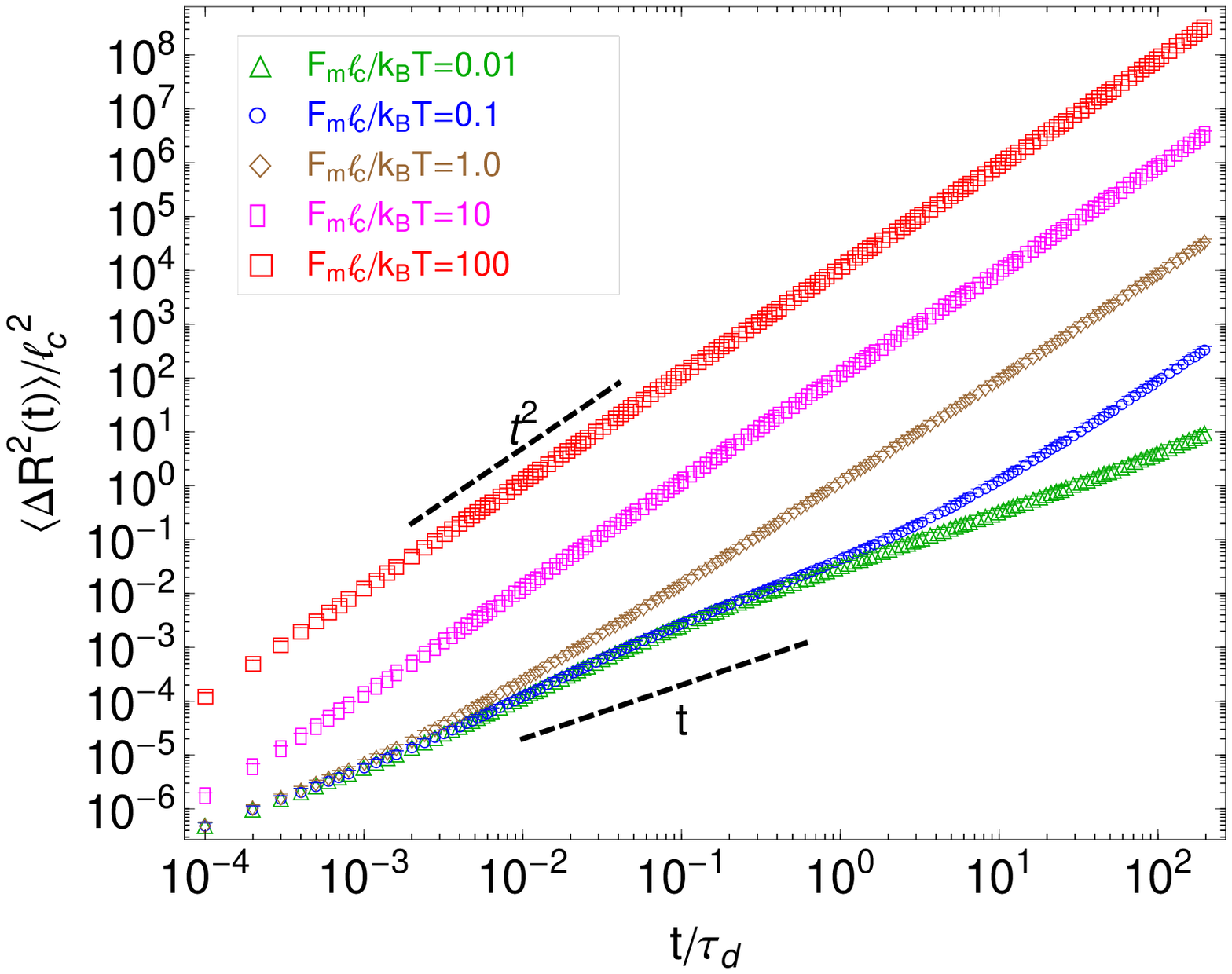}
\put (18,80) {\small (A)}
\end{overpic}\hspace{3mm}
\begin{overpic}[width=0.48\linewidth]{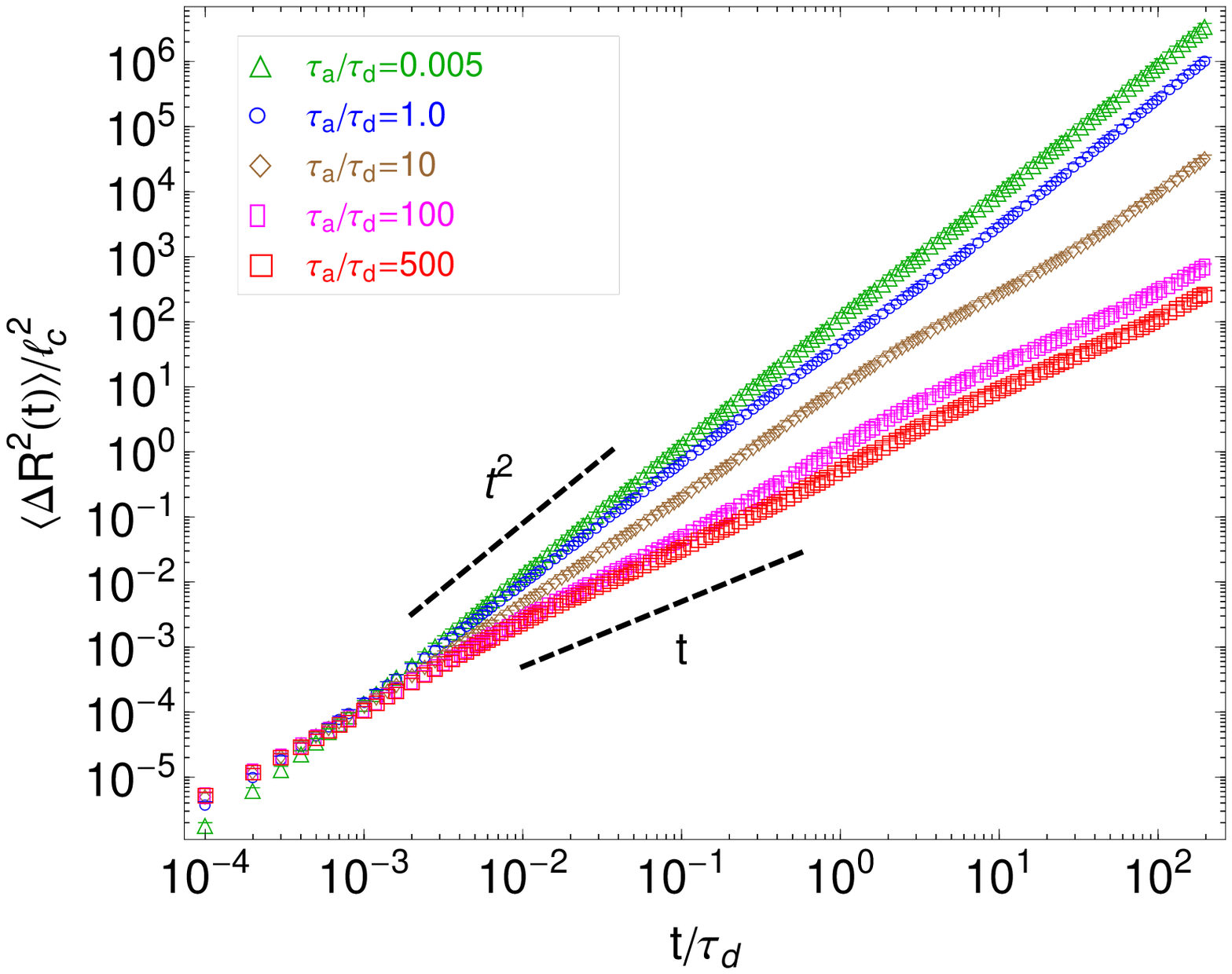}
\put (18,80) {\small (B)}
\end{overpic}
\caption{Steady state mean squared displacement of a
filament in the active single-chain model with $N=4$
as a function of A) the ratio between motor and  Brownian forces with 
$\tau_a/\tau_d=0.005$, and B) the ratio between attachment time
and detachment time of the motor clusters for $\dfrac{F_m \ell_c}{k_{\rm B}T}=10.0$. 
The other parameters used in these calculations were set to, 
$\dfrac{\zeta_a \ell_c^2}{\tau_d k_{\rm B}T}=192.1$, $\zeta_d/\zeta_a=0.1$, 
$\ell_p/\ell_c=10$.}
\label{fig3}
\end{figure}

In the active single-chain model  the mass transport of filaments 
inside and active gel can be quantified by the mean-squared displacement 
of a probe filament's center of mass, $\langle \Delta R_c^2(t) \rangle_{\rm st}= 
\langle[r_{\rm cm}(t)-r_{\rm cm}(0)]^2\rangle_{\rm st}$. 
Figure \ref{fig3}A shows the $\langle \Delta R^2 (t)\rangle_{\rm st}$ 
of a filament with four motor attachment sites, $N=4$, 
for different ratios of motor to Brownian forces, ranging from
a highly active gel, $\dfrac{F_m \ell_c}{k_{\rm B}T}=100$, to a mostly 
passive gel, $\dfrac{F_m \ell_c}{k_{\rm B}T}=0.01$. 
For this lowest ratio between motor and Brownian forces 
the $\langle \Delta R^2 (t) \rangle_{\rm st}$ exhibits 
diffusive behavior, $\sim t^{n\sim1}$, at all the time scales
sampled. For $\dfrac{F_m \ell_c}{k_{\rm B}T}=0.1$ apparent  
diffusive behavior, is also observed up to $t\lesssim \tau_d$ 
but super-diffusive behavior ($\sim t^{n>1}$) can be seen at longer times, 
$t>10\tau_d$. When $\dfrac{F_m \ell_c}{k_{\rm B}T}=1$ the 
diffusive region becomes much shorter and super-diffusive 
mass transport can be observed at time scales as short as $t\sim10^{-2}\tau_d$.
For gels with ratios of motor to Brownian forces higher than one the 
$\langle \Delta R^2 (t) \rangle_{\rm st}$ behaves ballistic, $\sim t^{2}$, at all 
the lag times sampled. The results show that,
as expected, a probe filament will \textcolor{black}{undergo diffusive motion} 
in the gel due to Brownian forces, as the magnitude of motor-generated forces 
becomes larger their pulling on the filament
becomes increasingly important and at long times they are able to 
effectively accelerate the motion of the filament's center of mass. For low 
ratios between motor and Brownian forces the latter are still
able to counteract, at short times, the directional pulling of the motors
with the Brownian motion of the filament segments. 
For large ratios between motor and Brownian forces 
the latter can not longer counteract the pulling done by the motor forces 
at any of the observable time scales.

Figure \ref{fig3}B  shows the $\langle \Delta R^2 (t)\rangle_{\rm st}$ for different 
ratios of motor detachment time, $\tau_d$, to attachment time, $\tau_a$, 
for a fixed value of the ratio between motor and Brownian forces,
$\dfrac{F_m \ell_c}{k_{\rm B}T}=10$. For the two highest $\tau_a/\tau_d$ ratios,
500 and 100, the behavior of the $\langle \Delta R^2 (t)\rangle_{\rm st}$ is 
ballistic at short time scales, $t\lesssim10^{-2}\tau_d$, and diffusive 
at longer time scales. This indicates that for these cases the motor attachment
times, $\tau_a$, are too long compared to the relaxation times of the filament segments.
Since the filaments segments have enough time to relax
between motor attachment events the motor proteins are not able to maintain 
the acceleration of the center of mass of the filaments at long times.
For the lowest $\tau_a/\tau_d$ ratios, 1.0 and 0.005, no significant local
relaxation can occur in the filament segments between motor attachment events and 
the behavior of $\langle \Delta R^2 (t)\rangle_{\rm st}$ is super-diffusive at all the 
observable lag times. These results indicate that even for a very active gel, where motor
forces dominate over Brownian forces, the mass transport of filaments will only
exhibit super-diffusive behavior at low $\tau_a/\tau_d$ ratios, when the motors
spend enough time pulling on the filaments and are able to accelerate their motion. 
These dependence of the $\langle \Delta R^2 (t)\rangle_{\rm st}$ 
on $\tau_a/\tau_d$ is in agreement with what was observed in the 
active single-chain model that did not include Brownian forces \citep{cordoba2015role}. 
In that model as well for gels with weak motor attachment, 
large $\tau_a/\tau_d$ ratios, the motion of the filaments has two well-defined regions. 
At short time scales, shorter than $\tau_d$, the $\langle \Delta R^2 (t)\rangle_{\rm st}$ 
goes as $\sim t^2$. While for larger time scales the behavior 
becomes diffusive. For strong attachment cases, 
small $\tau_a/\tau_d$ ratios, the $\langle \Delta R^2 (t)\rangle_{\rm st}$
of the model without Brownian forces also exhibited ballistic behavior at all time scales.

\subsection{Dynamic Modulus}\label{DM}

To test the validity of FDT in active gels Mizuno et al. \cite{mizuno2007nonequilibrium} 
compared the complex compliance of actomyosin networks measured with 
driven and passive probe particle microrheology. 
In the driven experiment an optical trap is used to apply a small-amplitude 
oscillatory force, $\bm{f}^{\rm trap}$, with a frequency $\omega$  to the probe
particle. The complex compliance, $\alpha(\omega)$, is obtained form the relation 
$\bm{r}_{\rm b}(\omega)= \alpha(\omega) \bm{f}^{\rm trap}$,
where $\bm{r}_{\rm b}(\omega)$ is the measured bead position at steady state.
In a passive microrheology experiment no external force is applied to the bead
or a static harmonic trap is used to hold the bead near its steady state position and
the imaginary part of $\alpha(\omega)$ is obtained using the FDT,
$\alpha''(\omega)=\dfrac{\omega}{6 k_{\rm B} T}C(\omega)$.
Where $C(\omega)=\int_{-\infty}^\infty\langle \bm{r}_{\rm b}(t) \cdot \bm{r}_{\rm b}(0)
\rangle e^{-i\omega t} dt$ is the autocorrelation function of the bead position.
Note also that, the complex compliance of the probe
particle can be related to the creep compliance of the material by using
a generalized Stokes relation, $J''(\omega)=6\pi R\alpha''(\omega)$. 
Before addition of ATP the complex compliance of the actomyosin 
networks obtained with the passive and driven microrheology experiments 
have been observed to agree. When the gels are activated with ATP
a discrepancy is observed between the complex compliance obtained from the 
passive and driven microrheology experiments. This discrepancy appears at 
frequencies below 10 Hz and becomes larger with decreasing frequency.

\begin{figure}[h t]
\vspace{5mm}
\begin{overpic}[width=0.48\linewidth]{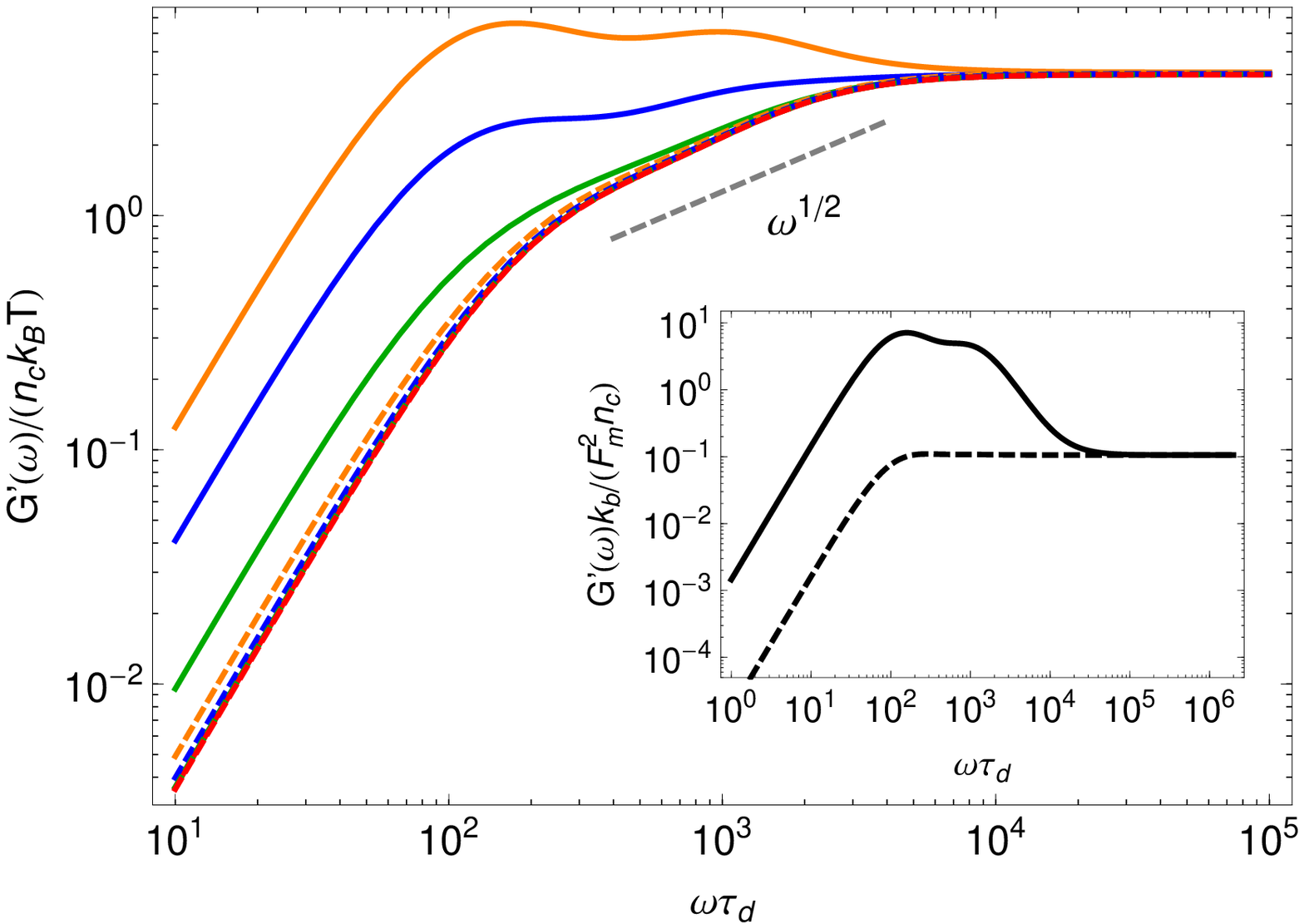}
\put (18,72) {\small (A)}
\end{overpic}\hspace{5mm}
\begin{overpic}[width=0.48\linewidth]{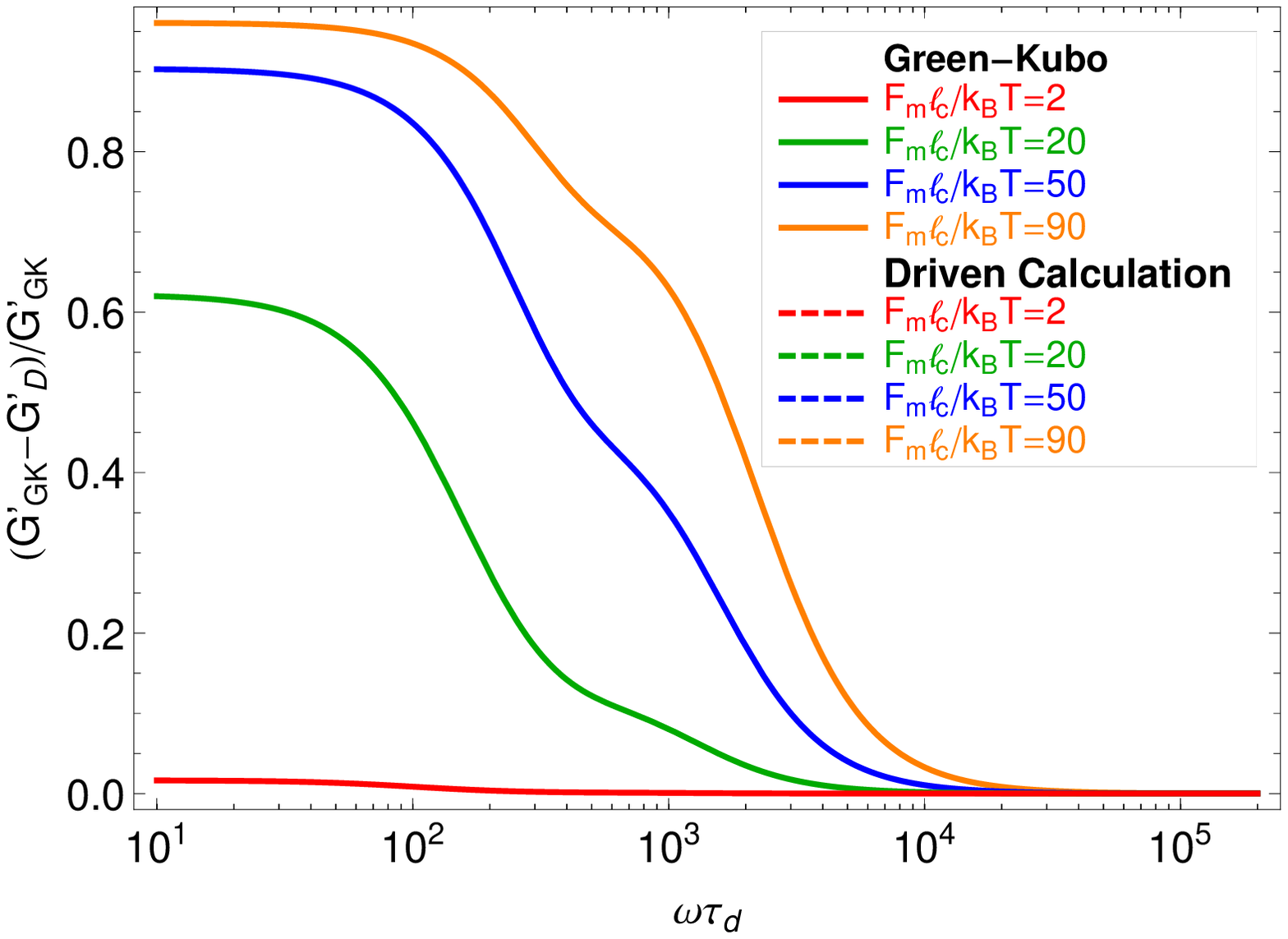}
\put (18,73) {\small (B)}
\end{overpic}
\caption{
(A) Storage moduli of the active single-chain dumbbell model, $N=2$, and
Dirac delta motor force distribution, $p(F)=\delta(F-F_m)$ for different values
of the ratio between motor and Brownian forces. The other parameters used in these 
calculations were set to, $\dfrac{\zeta_a \ell_c^2}{\tau_d k_{\rm B}T}=192.1$, 
$\zeta_d/\zeta_a=0.1$, $\tau_a/\tau_d=0.005$, $\ell_p/\ell_c=10$. The inset shows 
the storage modulus for the active single-chain dumbbell model without Brownian forces
with $p(F)=\delta(F-F_m)$, $\zeta_a/k_b=0.02$, $\zeta_d/\zeta_a=0.1$ and 
$\tau_a/\tau_d=0.005$. The continuous lines correspond to Green-Kubo
calculations and the dashed lines to driven calculations. 
(B) Check of the fluctuation dissipation theorem in the systems shown in part (A).}\label{fig1}
\end{figure}

In previous work \citep{Cordoba2014, cordoba2015role} it was shown that
the active single-chain model without Brownian forces is able
to describe the violation of the FDT observed in microrheology experiments
in active gels. Here the active single-chain model with Brownian forces 
will be tested in the same way. To do this the relaxation modulus of the active gel is
calculated by two different methods. In one method no 
external strain is applied, $\dot{\epsilon}(t)=0$ in eq.(\ref{CKeq}), 
and the Green-Kubo formula is used to calculate the  relaxation modulus of the material 
from the autocorrelation function of stress at steady state \cite{cordoba2015role}, 
$G_{\rm GK}(t)=\dfrac{1}{n_c k_{\rm B} T}\left \langle \sigma(t) 
\sigma(0) \right \rangle_{\rm{st}}$. For a mean-field single-chain model, like the 
one employed here, the macroscopic stress, $\sigma$, is related to the tension on the filaments 
by $\sigma=-n_c\sum_{i=1}^{N-1} f_i r_i$ where $f_i$ is the tension on 
a filament segment and $n_c$ is the number of filaments per unit volume 
\cite{indei2010linear, bird1978dynamics, hernandez2003brownian}. 
For calculations with Fraenkel springs the macroscopic stress simplifies to 
$\sigma=n_c k_b \sum_{i=1}^{N-1} r_i^2$. The non-equilibrium steady state dynamic 
modulus of the active gel can then be obtained 
by taking the one-sided Fourier transform of the relaxation modulus,
$G^*(\omega)=i\omega\bar{\mathcal{F}}\{G(t)\}=G'(\omega)+i G''(\omega)$.
Where $\bar{\mathcal{F}}\{G(t)\}:=\int_0^\infty G(t) e^{-i 
\omega t}dt$ is the one-sided Fourier transform; 
$G'(\omega)$ is the storage modulus and $G''(\omega)$ 
is the loss modulus. In a second calculation, 
the dynamic modulus is obtained from the stress response to an externally 
applied small-amplitude oscillatory strain or from a small step strain. 
For the dumbbell version of the model, $N=2$, and a Dirac delta motor force 
distribution, $p(F)=\delta(F-F_m)$, these two calculations
can be carried out analytically. The details of the analytical procedure to derive 
the dynamic modulus from these two types
of calculations has been presented elsewhere \citep{Cordoba2014}.

Figure \ref{fig1}A shows the storage modulus for the dumbbell version
of the active single-chain model with a Dirac delta motor force distribution
for different ratios of motor to Brownian forces. Note that for the
dumbbell case, the gel is formed by filaments of length $\ell_c$ with only two 
motor attachment sites per filament. Again, to relate the model predictions
to experimental observations it will be considered that the
the ratio of motor to Brownian forces is proportional to the 
concentration of ATP in the active gel. Higher ATP concentrations yield 
higher motor to Brownian forces ratios while ATP-depleted gels have lower motor
 to Brownian forces ratios. Figure \ref{fig1}B shows the same type of 
 frequency dependent violation of the FDT
that was previously observed with the model that did not include
Brownian forces. The modulus obtained from the
Green-Kubo formula and the modulus obtained from the driven calculation agree 
at high frequency, but diverge at low frequencies. Moreover, this discrepancy at 
low frequencies becomes larger with increasing ratio between motor and Brownian forces. 
For the higher ratios between motor and Brownian forces a maximum appears 
in the storage moduli obtained from the Green-Kubo formula at frequencies
around $10^2/\tau_d$. This feature of the dynamic modulus of active gels  
had already been observed before in the version of
the model without Brownian forces \citep{Cordoba2014} and is also seen here 
for large ratios of motor to Brownian forces.
For comparison, the storage modulus of the active single-chain model
without Brownian forces is shown in the inset of Figure \ref{fig1}A.
Note that this modulus has been calculated with parameters that are equivalent
to the ones used for the model with Brownian forces. That is,
$\zeta_a/k_b=0.02$, $\zeta_d/\zeta_a=0.1$ and $\tau_a/\tau_d=0.005$.
Also important is that in the limit of very 
small ratios of motor to Brownian forces the modulus from the passive calculation
agrees exactly with the modulus from the active calculation at all frequencies and the FDT 
is recovered (Figure \ref{fig1}B), as expected for a passive polymer network. 

\begin{figure}[h t]
\vspace{5mm}
\begin{overpic}[width=0.48\linewidth]{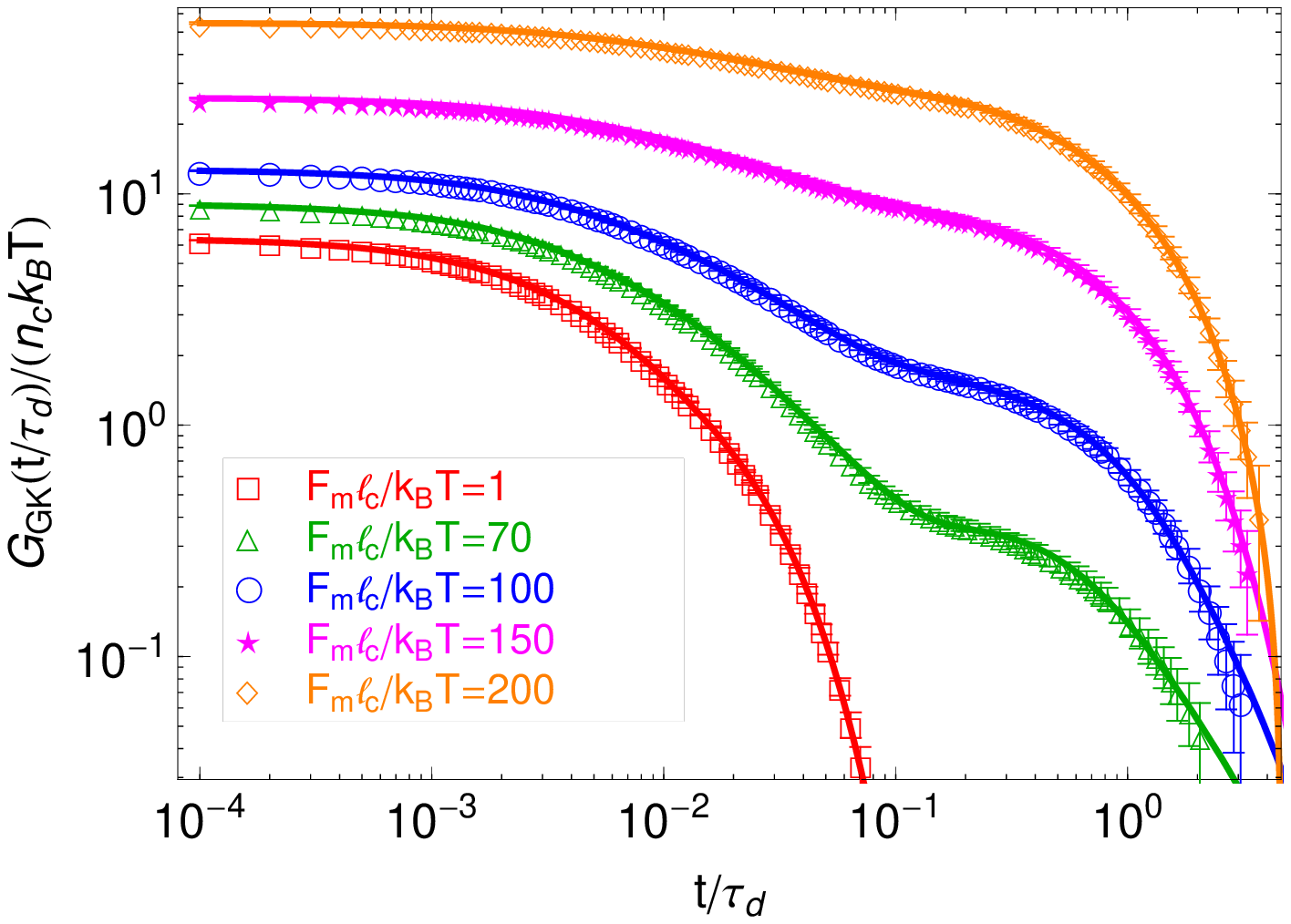}
\put (18,72) {\small (A)}
\end{overpic}\hspace{3mm}
\begin{overpic}[width=0.48\linewidth]{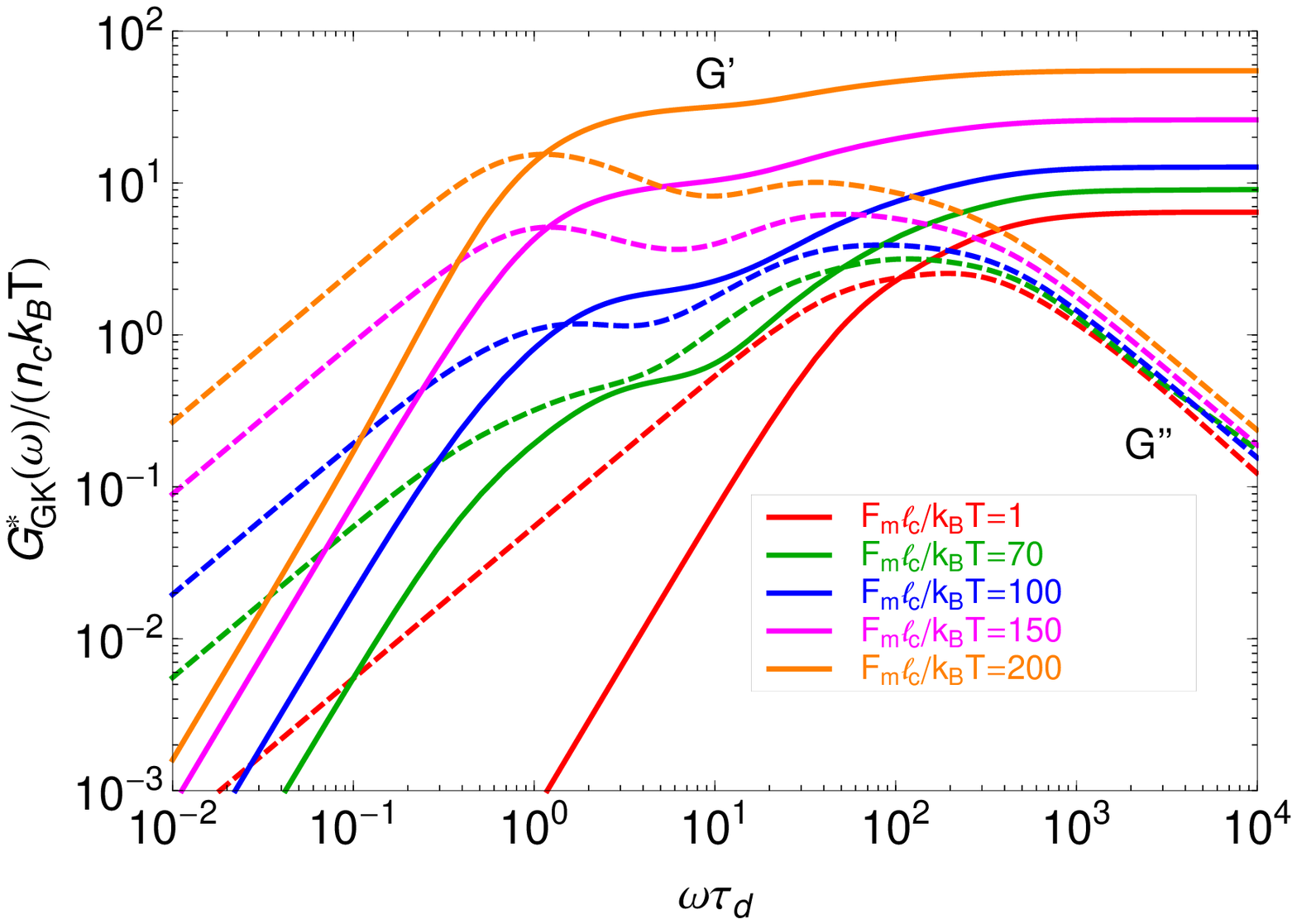}
\put (18,72) {\small (B)}
\end{overpic}
\caption{
(A) Relaxation moduli, calculated with the Green-Kubo formula,  
of the  active single-chain model with $N=4$ and for different values of the ratio 
between motor and  Brownian forces. 
\textcolor{black}{The symbols are simulation results and the lines are fits to a multimode
Maxwell model used to transfer the information to the frequency domain.}
The other parameters used in these 
calculations were set to, $\dfrac{\zeta_a \ell_c^2}{\tau_d k_{\rm B}T}=192.1$, 
$\zeta_d/\zeta_a=0.1$, $\tau_a/\tau_d=0.005$, $\ell_p/\ell_c=10$. 
Here and below the error bars on the autocorrelation of stress, $G_{\rm GK}(t)$,
were calculated using the ``blocking transformations'' method, which 
accounts for the correlation in the time-series data \citep{flyvbjerg1989error}.
(B) Dynamic modulus of the active single-chain model calculated from the 
relaxation moduli shown in part (A). 
\textcolor{black}{The continuous lines correspond to the storage moduli, $G'$, and 
the dashed lines to the loss moduli, $G''$.}}
\label{fig2}
\end{figure}

The inclusion of Brownian forces in the model also introduces additional features in the shape 
of the storage modulus that are not observed in the dumbbell version of the model without
Brownian forces. In particular, a scaling of $G'\sim \omega^{0.5}$ appears at 
intermediate frequencies, $2\times10^2\lesssim \omega\tau_d \lesssim 10^4$ ,
for the lower motor to Brownian forces ratios. 
This stress relaxation mechanisms is of the Rouse type and is also observed in temporary 
networks formed by associating polymers \cite{indei2010linear, indei2012effects}, 
where it is called associative Rouse behavior, to distinguish it from another Rouse 
relaxation region observed at high frequencies in those gels. 
For the active single-chain model with Brownian forces, the Rouse behavior appears 
in the frequency region bounded by $\tau_{r,a}=\zeta_a/k_b=0.02\tau_d$, 
and $\tau_{r,d}=\zeta_d/k_b=0.002\tau_d$. 
In the active single-chain model without Brownian forces this type of behavior is not 
observed for dumbbells (see inset of Figure 3A) and only appeared for longer filaments
where the longest relaxation time of the gel becomes
significantly larger than $\tau_d$. Here, however this relaxation
mode is observed at time scales between three and two orders of 
magnitude smaller than $\tau_d$. This indicates that Brownian forces introduce 
faster relaxation dynamics that were not present in the model without Brownian
forces. However final conclusions should not be drawn from the dynamics 
of the dumbbell versions of the model since these are unrealistically fast. More realistic
systems, that require numerical solutions of the model, will be addressed below. 
Still the analytic calculations that can be performed with the dumbbell model provide
a good reference point for the more complex numerical simulations.

In Figure \ref{fig2}A the relaxation moduli for gels formed by
filaments of length $\ell_t=3\ell_c$, or equivalently filaments with $N=4$,
and for five different ratios of motor to Brownian forces are shown. 
In Figure \ref{fig2}B the corresponding dynamic moduli are presented. 
For these calculations the motor forces were
generated from a gamma distribution with scale parameter
$\beta=0.42 F_m$. Note that when the mean 
motor stall force, $F_m$, is varied the shape parameter of the
gamma distribution, $\alpha=F_m/\beta=2.4$, remains constant.
These shape and scale parameters were obtained by fitting a
a gamma probability distribution to a motor force
distribution measured in actomyosin bundles \citep{thoresen2011reconstitution}. 
As expected, with these longer filaments and the more realistic motor force distribution
a broader spectrum of relaxation times
can be observed compared to the results for the dumbbell filaments.  For the lowest 
ratio between motor and Brownian forces, $\dfrac{F_m \ell_c}{k_{\rm B}T}=1.0$,
the dynamic modulus has the shape typically observed in temporary polymeric networks.
A cross-over between $G'$ and $G''$, which is associated with the longest relaxation time of
the network is observed at frequencies around $10^2/\tau_d$. Below this frequencies 
the terminal zone of stress relaxation is observed with 
$G'\sim\omega^2$ and $G''\sim\omega$. At high frequencies
$G'$ exhibits a plateau region.
 
As the ratio of motor to Brownian forces is increased the magnitude and 
shape of $G_{\rm GK}(t)$ and therefore of $G^*_{\rm GK}$ change 
significantly. These changes are more pronounced at
low frequencies, below $10^2/\tau_d$, where new or additional 
local maxima appear both in $G'$ and $G''$. 
For a ratio of motor to Brownian forces of
$\dfrac{F_m \ell_c}{k_{\rm B}T}=70$ new features in 
$G'$ and $G''$ are observed around 
$1/\tau_d\lesssim\omega\lesssim 10^2/\tau_d$. Moreover 
the crossover between $G'$ and $G''$ is slightly displaced to lower frequencies. 
As the ratio between motor and Brownian forces is increased
the overall magnitude of $G^*_{\rm GK}$ increases further,
the size of the low-frequency peaks in $G''$ also increases and
the crossover frequency between $G'$ and $G''$ moves 
to even lower frequencies. In particular, this last observation
indicates that a $200$ fold increase in the ratio between 
motor and Brownian forces will increase the longest relaxation time
of the gel by two orders of magnitude. Note that in the model
with Brownian forces $G''$ always exhibits two maxima, a low frequency one around
the crossover and one at higher frequencies. The one at higher 
frequencies was not observed in the model without Brownian forces.

\begin{figure}[h t]
\vspace{5mm}
\begin{overpic}[width=0.48\linewidth]{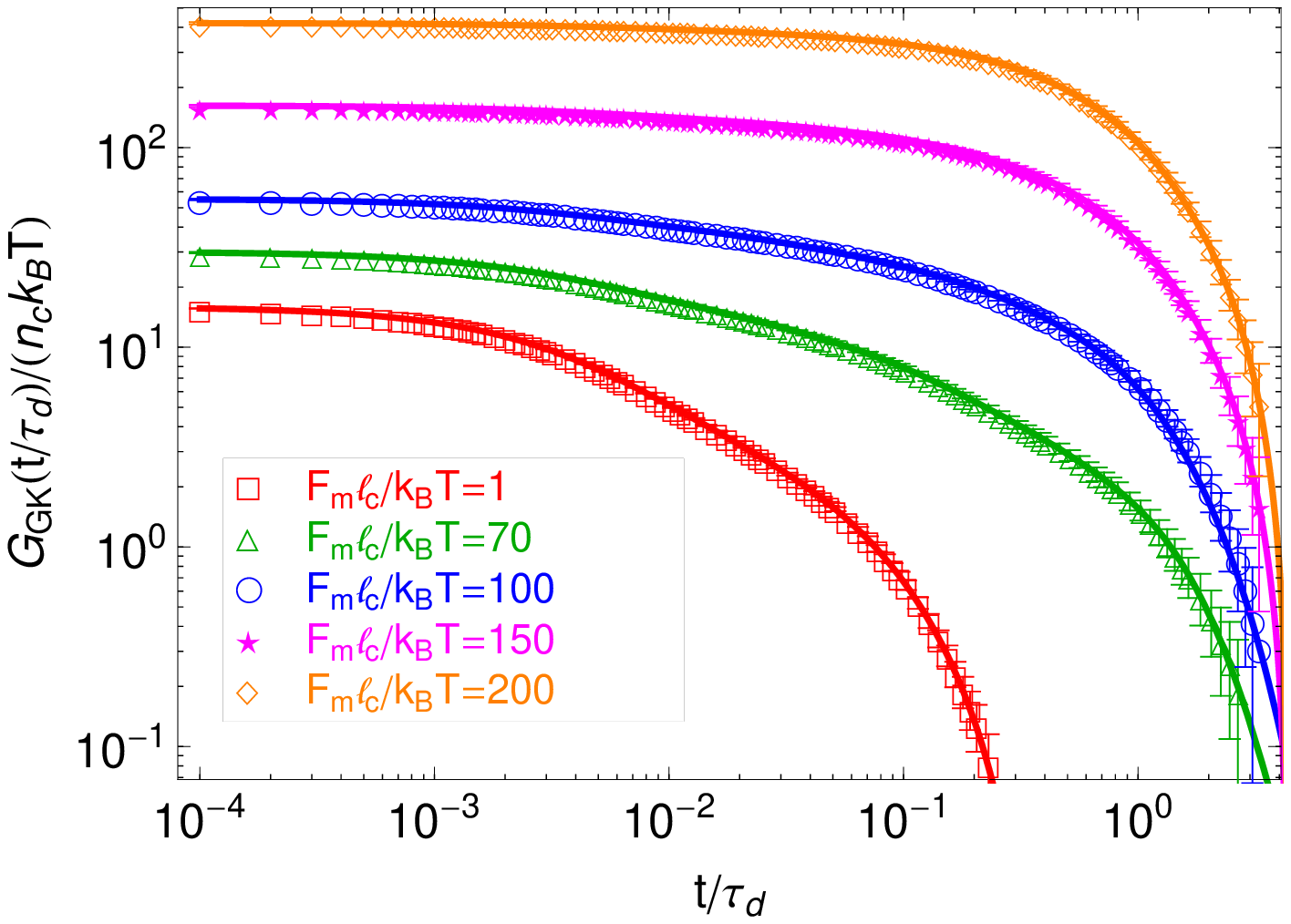}
\put (18,72) {\small (A)}
\end{overpic}\hspace{3mm}
\begin{overpic}[width=0.48\linewidth]{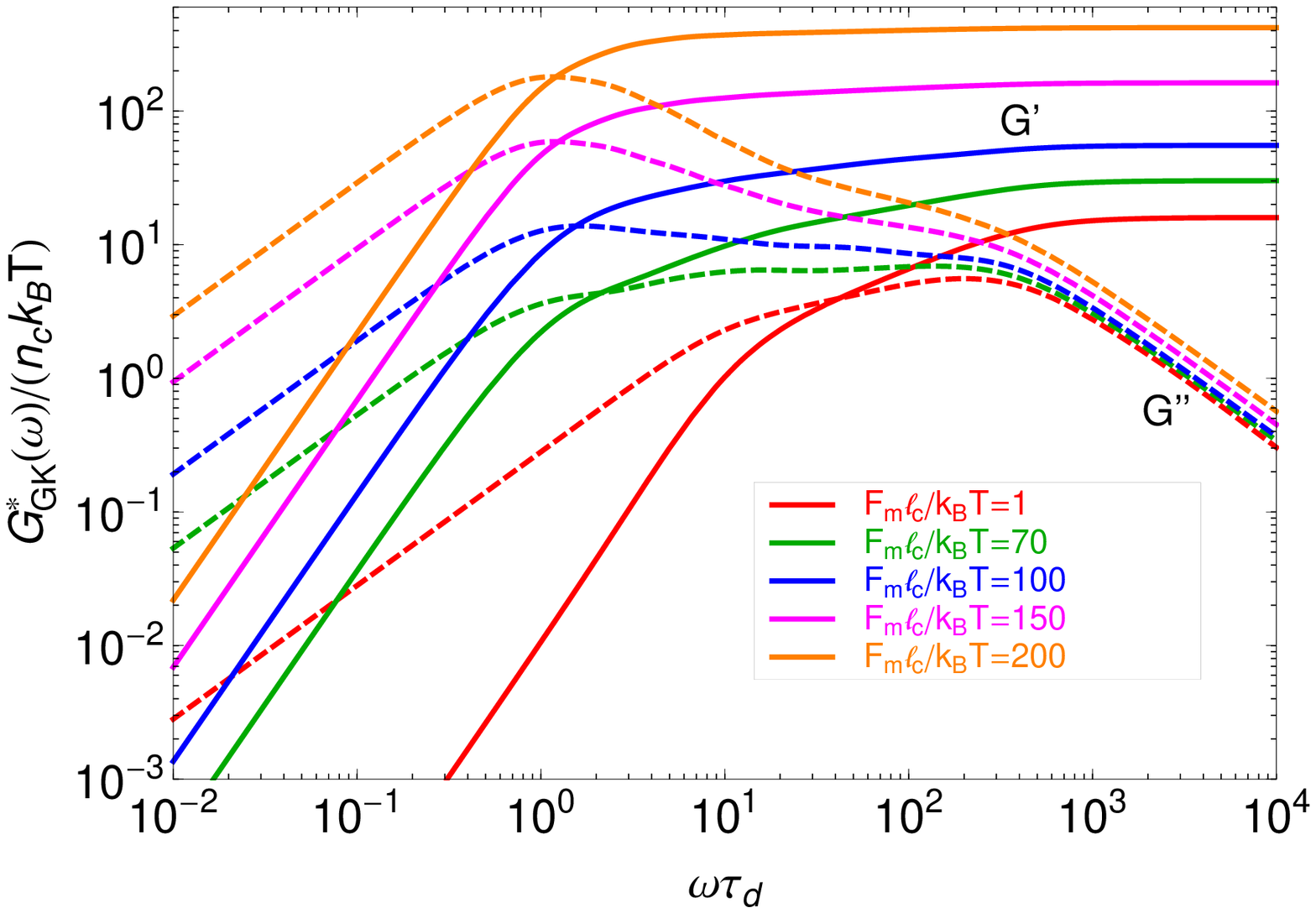}
\put (18,72) {\small (B)}
\end{overpic}
\caption{
(A) Relaxation moduli, calculated with the Green-Kubo formula,  
of the  active single-chain model with $N=8$ and for different values of the ratio 
between motor and  Brownian forces. 
\textcolor{black}{The symbols are simulation results and the lines are fits to a multimode 
Maxwell model used to transfer the information to the frequency domain.}
The other parameters used in these 
calculations were set to, $\dfrac{\zeta_a \ell_c^2}{\tau_d k_{\rm B}T}=192.1$, 
$\zeta_d/\zeta_a=0.1$, $\tau_a/\tau_d=0.005$, $\ell_p/\ell_c=10$. 
(B) Dynamic modulus of the active single-chain model calculated from the 
relaxation moduli shown in part (A). 
\textcolor{black}{The continuous lines correspond to the storage moduli, $G'$, and 
the dashed lines to the loss moduli, $G''$.}}
\label{fig4}
\end{figure}

Figure \ref{fig4} illustrates the effect on $G_{\rm GK}(t)$ and $G^*_{\rm GK}$
when the filament length is increased to $\ell_t=7\ell_c$, or equivalently 
filaments with $N=8$. The gel formed by these even longer filaments and 
with $\dfrac{F_m \ell_c}{k_{\rm B}T}=1$  exhibits a longer power law relaxation mode, 
$G'\sim\omega^{1/2}$ at frequencies near the crossover between $G'$ and $G''$. 
Moreover, this crossover is also shifted to lower frequencies, as expected for 
gels formed by higher molecular weight filaments. 
As the ratio between motor and Brownian forces is increased the magnitudes 
of $G_{\rm GK}(t)$ and $G^*_{\rm GK}$ also increase significantly. Also,
the longest relaxation time increases by two orders of magnitude when
the ratio between motor forces and Brownian forces is increased by a factor 
of 200. The low frequency maxima are still present for both $G'$ and $G''$.
However, for these longer filaments the maximum in $G''$ 
that appeared at higher frequencies, $\omega\sim 10^2/\tau_d$, 
is very small which seems to indicate that it will tend to disappear as the 
filaments are made longer.

\begin{figure}[h t]
\vspace{5mm}
\begin{overpic}[width=0.48\linewidth]{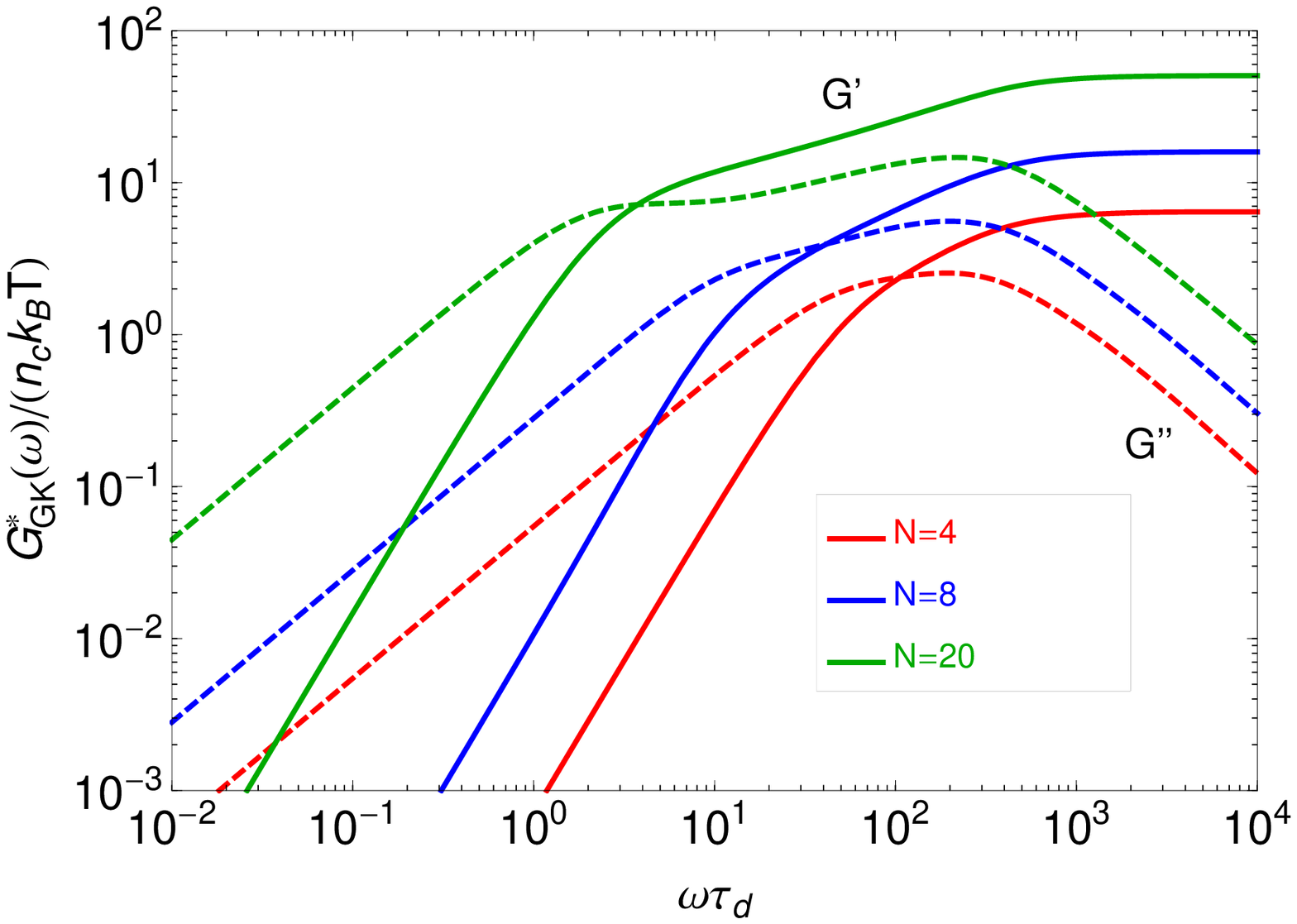}
\put (18,72) {\small (A)}
\end{overpic}\hspace{3mm}
\begin{overpic}[width=0.48\linewidth]{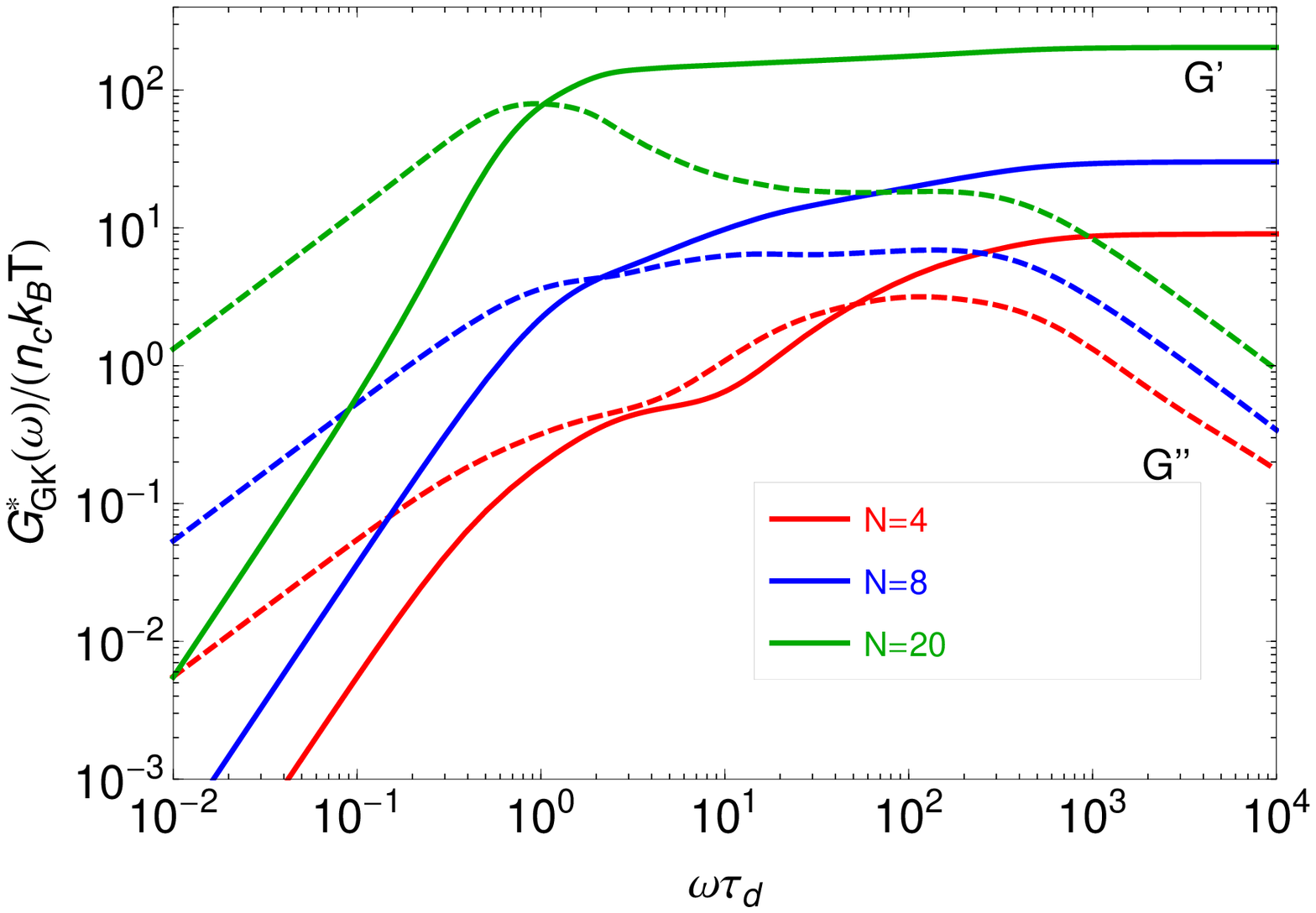}
\put (18,72) {\small (B)}
\end{overpic}
\caption{
(A) Dynamic moduli of the active single-chain model  
with $\dfrac{F_m \ell_c}{k_{\rm B}T}=1$
calculated from Green-Kubo simulations for different values
of the filament length, $\ell_t=(N-1)\ell_c$.
(B) Dynamic moduli of the active single-chain model 
with $\dfrac{F_m \ell_c}{k_{\rm B}T}=70$
calculated from Green-Kubo simulations for different values
of the filament length, $\ell_t=(N-1)\ell_c$.
The other parameters used in these calculations were set to, 
$\dfrac{\zeta_a \ell_c^2}{\tau_d k_{\rm B}T}=192.1$, $\zeta_d/\zeta_a=0.1$, 
$\tau_a/\tau_d=0.005$, $\ell_p/\ell_c=10$. 
\textcolor{black}{The continuous lines correspond to the storage moduli, $G'$, and 
the dashed lines to the loss moduli, $G''$.}}
\label{fig5}
\end{figure}

\textcolor{black}{Note that the associative Rouse relaxation 
mode, $G'\sim\omega^{1/2}$, observed in the calculations with the active single-chain model 
is a result of the simplified description of the polymeric filaments.
In the active single-chain model the filament segments are described by Fraenkel springs. 
However, real semiflexible filaments have finite-extensibility and the tension in 
a particular segment is strongly coupled to the orientation of the other segments in 
the filament. The effect of finite-extensibility in the rheology of active gels 
was previously addressed using the active single-chain model without Brownian 
forces \cite{cordoba2015role}. Moreover, a more accurate bead-spring chain discretization
of the semiflexible filaments should also include bending potentials between the springs which
introduce correlations between the orientations of the filament segments 
\cite{underhill2006alternative, koslover2013discretizing}.}

To illustrate how the stress relaxation dynamics in active gels 
behave for even longer filaments Figure \ref{fig5} shows a 
comparisons between the dynamic moduli of filaments with $N=4$, $N=8$ and $N=20$
for ratios of motor to Brownian forces,  $\dfrac{F_m \ell_c}{k_{\rm B}T}=1$ 
and $\dfrac{F_m \ell_c}{k_{\rm B}T}=70$. Note that for the gels with the lower
ratio between motor and Brownian forces increasing the filament length
by a factor of about five increases the longest relaxation time
by two orders of magnitude. Moreover, the breadth of the 
Rouse-type low-frequency relaxation mechanism, $G'\sim\omega^{1/2}$,
increases from one decade to three decades of frequency when
the filament length increases by a factor of about five. 
Also as expected, gels formed by longer filaments exhibit 
dynamic moduli with overall higher magnitude. For the gels
with $\dfrac{F_m \ell_c}{k_{\rm B}T}=70$ the crossover between 
$G'$ and $G''$ moves to lower frequencies and the overall
magnitude of $G^*_{\rm GK}$ increases as $N$ is increased.
Also, for these gels with larger motor forces the low frequency 
peaks in $G'$ and $G''$ become larger as $N$ is increased.
For instance, for $N=4$ the low frequency peak that appears
in $G''$ at $\omega\sim 1/\tau_d$ is small compared to
the higher frequency peak that appears at $\omega\sim 10^2/\tau_d$.
For $N=8$ these two peaks have similar magnitude and 
$G''$ exhibits a long plateau between 
$1/\tau_d\lesssim\omega\lesssim 10^2/\tau_d$. Then for 
$N=20$ the lower frequency peak in $G''$ becomes much larger
than the higher frequency peak and a clear maximum also
appears in $G'$ at frequencies around the crossover between $G'$ and $G''$.
Note that the experimental counterpart of the Gree-Kubo simulations discussed 
so far are passive microrheology experiments. In the next subsection simulation
results that are comparable to driven rheological experiments will be discussed.

\subsection{Comparisons to microrheology experiments in actomyosin gels}\label{PRED}

\begin{figure}[h t]
\vspace{5mm}
\begin{overpic}[width=0.48\linewidth]{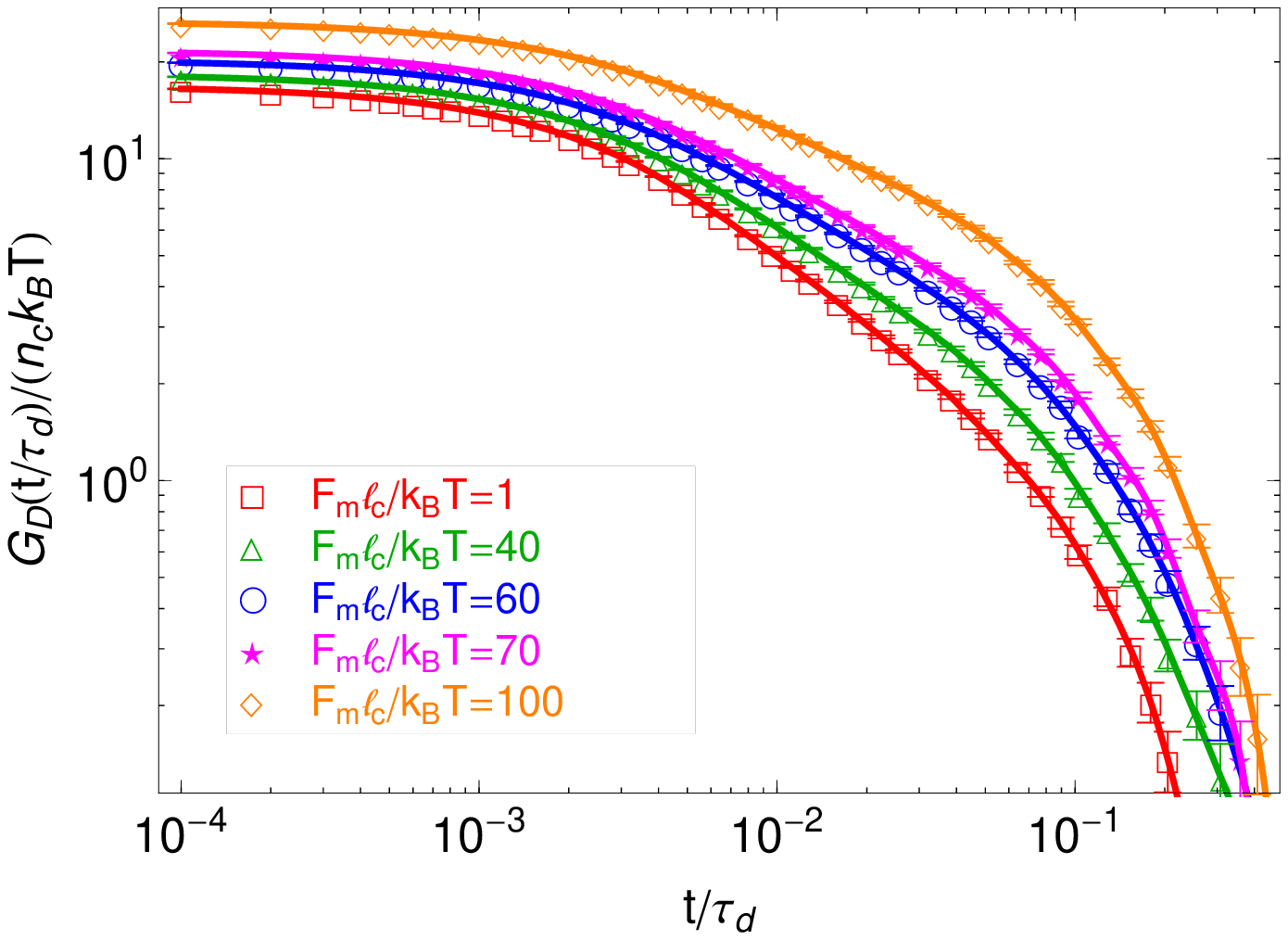}
\put (85,72) {\small (A)}
\end{overpic}
\begin{overpic}[width=0.48\linewidth]{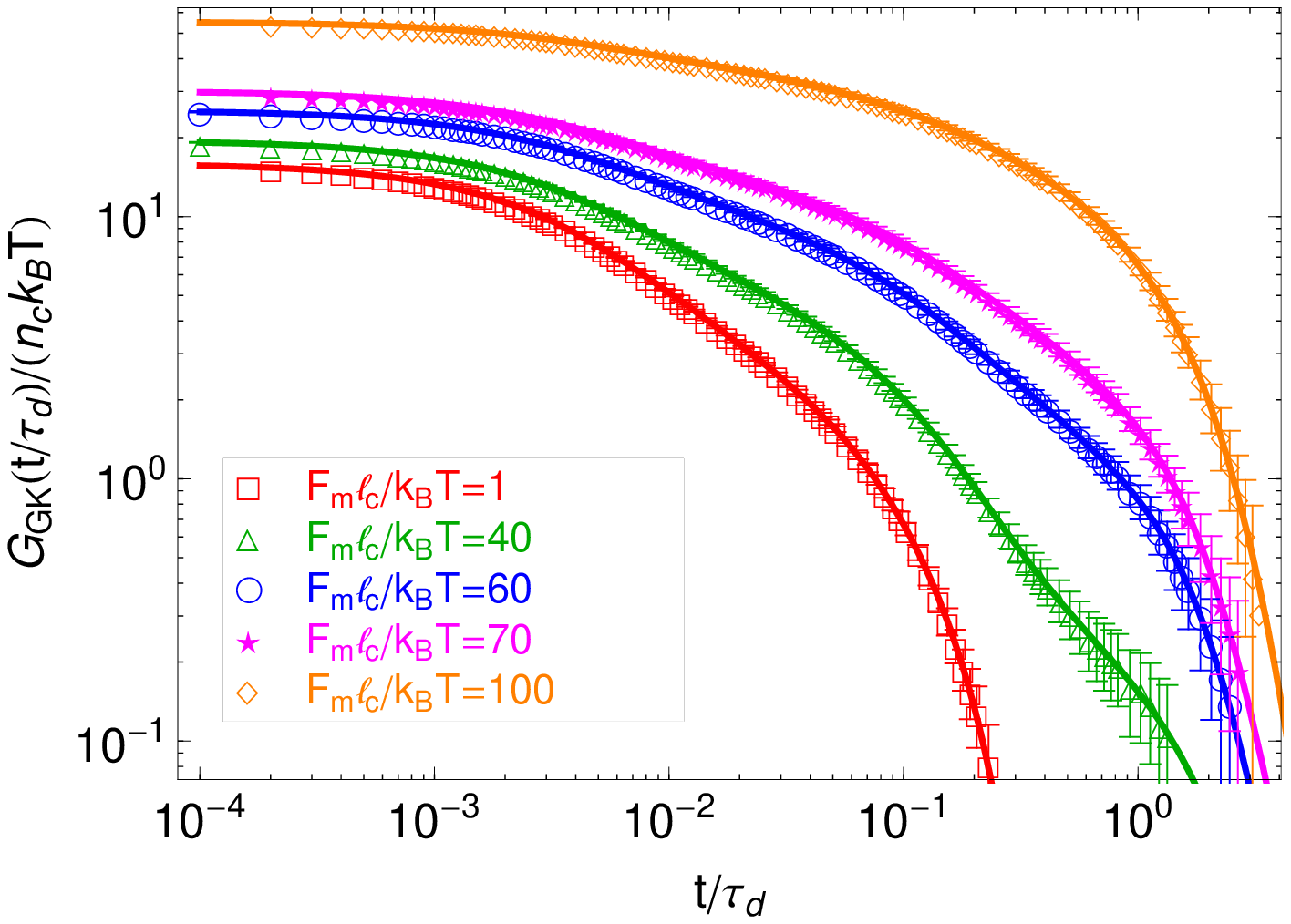}
\put (85,72) {\small (B)}
\end{overpic}\hspace{3mm}
\caption{
Relaxation modulus of the active single-chain model 
as a function of the ratio between motor-generated and Brownian forces.
\textcolor{black}{The symbols are simulation results and the lines are fits to a
multimode Maxwell model used to transfer the information to the frequency domain.}
For these calculations the parameters are set to, $N=8$,
$\dfrac{\zeta_a \ell_c^2}{\tau_d k_{\rm B}T}=192.1$, $\zeta_d/\zeta_a=0.1$, 
$\tau_a/\tau_d=0.005$, $\ell_p/\ell_c=10$.
(A) Calculated from driven step-strain simulations.
(B) Calculated from the steady-state 
autocorrelation function of stress using the Green-Kubo formula.}
\label{fig6}
\end{figure}

To compare the active single-chain model predictions
to driven microrheology experiments
the relaxation modulus obtained from the stress response to an 
externally applied small step-strain is calculated. To do this,
an ensemble of chains with dynamics given by 
eq.(\ref{CKeq}) was simulated. These simulations 
were started from an initial condition in which
all the strands are relaxed, $r_i=0$ for all $i$, and then the ensemble of chains
is allowed to reach steady state before applying a small step-strain of magnitude
$\epsilon_0$, at $t=t_{\rm st}$. Simulations are performed with progressively 
smaller values of $\epsilon_0$ to check for convergence to 
the linear response regime. It is assumed that on the time scale of interest for which $G(t)$ 
is calculated the step-strain applied at the boundaries propagates instantaneously through
the system. Therefore $r_i(t=t_{\rm st}+)=r_i(t=t_{\rm st}-)+\epsilon_0 r_i(t=t_{\rm st}-)$
for $i=1, 2, 3, ..., N-1$, where $\epsilon_0$ is the strain magnitude. 
Then the relaxation of stress back to its steady state value is followed. 
In this externally driven calculation the relaxation modulus is given by
$G_{\rm D}(t)=\dfrac{\langle \sigma(t) \rangle_{\rm st}}{\epsilon_0}$. 
The subscript D indicates that the modulus is obtained from an externally driven 
experiment (externally applied strain). To obtain the plots shown
here an average over an ensemble of $2.1\times10^5$ filaments is taken.
In Figure \ref{fig6}A the relaxation moduli as a function of the ratio between
motor and Brownian forces for gels with $N=8$ are shown. For comparison,
the relaxation moduli obtained from the steady state 
autocorrelation function of stress using the Green-Kubo formula are shown
in Figure \ref{fig6}B. In both types of calculations the calculated relaxation modulus
exhibits an overall increase in magnitude as the ratio between active and Brownian
forces is increased. Also, both calculations show a slowing down of the 
stress relaxation dynamics as the ratio of motor to Brownian forces is 
increased. However, both of these effects are
much more pronounced in $G_{\rm GK}(t)$ than in $G_{\rm D}(t)$.

\begin{figure}[h t]
\vspace{5mm}
\begin{overpic}[width=0.48\linewidth]{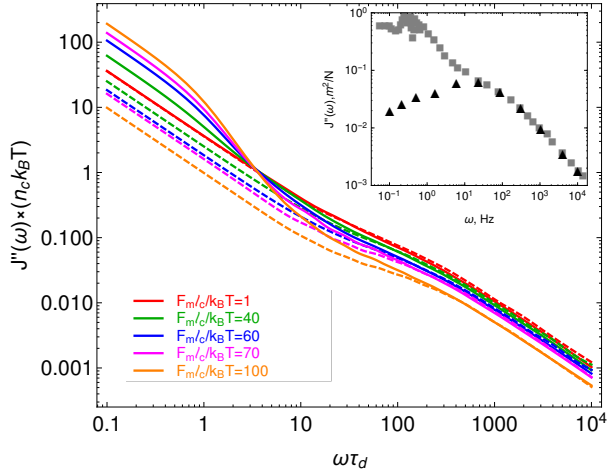}
\end{overpic}
\caption{ The Imaginary part of the creep compliance of an active gel
as a function of the ratio between motor-generated and 
Brownian forces calculated with the active single-chain 
model with Brownian forces. The dashed lines
are obtained from driven step-strain simulations, while 
the continuous lines are obtained from Green-Kubo simulations.
In all these calculations the model parameters are set to, $N=8$,
$\dfrac{\zeta_a \ell_c^2}{\tau_d k_{\rm B}T}=192.1$, $\zeta_d/\zeta_a=0.1$, 
and $\tau_a/\tau_d=0.005$ and $\ell_p/\ell_c=10$.
The inset shows the imaginary part of the creep compliance, 
$J''(\omega)=6\pi R\alpha''(\omega)$ observed in
driven microrheology (triangles) and passive 
microrheology (squares) in actomyosin gels 
\citep{mizuno2007nonequilibrium}, \textcolor{black}{where 
$R$ is the radius of the probe bead used in the microrheology experiments.
In these particular experiments the probe particles used had $R=2.5~\mu\text{m}$}.}
\label{fig8}
\end{figure}

To directly compare the active single-chain model predictions
to available microrheology experimental data in active gels the the complex creep 
compliance, $J^*(\omega)$, needs to be computed. To contrast the model
predictions to driven microrheology experiments the imaginary 
part of the creep compliance is calculated as 
$J_{\rm D}''(\omega)=\dfrac{{\rm Im}\{G^*_{\rm D}(\omega)\}}
{|G^*_{\rm D}(\omega)|^2}$. While to compare to passive microrheology experiments the 
imaginary part of the creep compliance \cite{levine2009mechanics, head2010nonlocal} is computed as  
$J_{\rm GK}''(\omega)=\dfrac{{\rm Im}\{G^*_{\rm GK}(\omega)\}}{|G^*_{\rm D}(\omega)|^2}$. 
Where $G^*_{\rm GK}(\omega)$ is the dynamic modulus 
obtained from a Green Kubo simulation and $G^*_{\rm D}(\omega)$ is the dynamic modulus 
obtained from a step-strain simulation. Figure \ref{fig8}
shows the resulting $J''(\omega)$ obtained with the active single chain model
for different values of the ratio between motor and Brownian forces. 
For comparison the inset shows the $J''(\omega)$ of actomyosin gels 
obtained from passive (squares) and driven (triangles) microrheology 
experiments \citep{mizuno2007nonequilibrium}. 
Note the characteristic frequency-dependent discrepancy 
between the $J''(\omega)$ obtained from passive and driven
microrheology experiments. Both types of experiments
agree at high frequencies but diverge at low frequencies, below 10 Hz.

At very low frequencies the model predictions
do not follow the shape of the $J''(\omega)$ observed in driven
microrheology experiments, while the calculated $J''$ increase with $\omega$ at
low frequencies the experimental data decays with 
$\omega$ at low frequencies. The shape of $J''$ observed in
the driven microrheology experiments is typical of polymeric gels with permanent 
cross-links. Therefore that particular discrepancy between the model predictions and 
the experimental data can be attributed to the presence of biotin cross-links 
in the actomyosin gels prepared by Mizuno et al \cite{mizuno2007nonequilibrium}.
Besides this rather obvious discrepancy the model predictions for 
large ratios between motor and Brownian forces describe
well the frequency dependence of $J''(\omega)$ observed in driven and passive
experiments in active actomysoin gels. The model 
predictions show that for high motor to Brownian forces ratios 
$J_{\rm D}''$ and $J_{\rm GK}''$ are equal at high frequencies but
the characteristic low frequency discrepancy appears around $\omega\sim10^2/\tau_d$.
For low  $\dfrac{F_m \ell_c}{k_{\rm B}T}$ the low frequency difference 
between $J_{\rm D}''$ and $J_{\rm GK}''$ disappears which indicates that the 
FDT becomes valid and the gel can be regarded as passive. 
Is also worth comparing the predictions of the active single-chain
model to the predictions of its counterpart without Brownian forces. 
This latter model also described correctly the low frequency discrepancy 
in the rheological response of active gels. However the breadth of the 
relaxation spectrum, especially at higher frequencies, was significantly more
narrow in the model without Brownian forces. Also, the passive
gel limit, where the FDT is recovered, was not attainable
in the model that did not include Brownian forces. Moreover, 
the wider relaxation spectrum of the active single-chain model with Brownian 
forces represents better the observations in rheological experiments 
in actomyosin gels. 

\textcolor{black}{
Note that for values of $\ell_c\sim2~\mu\text{m}$ and 
$\tau_d\sim200~\text{ms}$ typical of actomyosin gels \cite{lenz2012contractile}
the friction coefficients that have been used for the model predictions throughout this work are, 
$\zeta_a\sim0.039~\mu\rm{N\cdot s/m}$ and $\zeta_d\sim0.0039~\mu\rm{N\cdot s/m}$.
With these values, the Stokes relation can then be used
to obtain a rough estimate the effective size of the motor clusters in the active gel. 
Using the viscosity of water as the solvent, $\eta\sim 10^{-3}~\rm{Pa\cdot s}$,
this calculation yields a value of $\xi\equiv\dfrac{\zeta_a}{6 \pi \eta}\sim2~\mu$m 
which is consistent with the typical size of myosin motor clusters which
has been reported \cite{murrell2012f} to be around $1.5~\mu$m.}

\section{Conclusions}\label{CO}

The results presented here indicate that a competition between 
motor-generated forces and Brownian forces is important
in producing characteristic features observed in the mass 
transport and rheological properties of active gels.  
Previous versions of the active single-chain model neglected Brownian forces,
this simplification was assumed to be appropriate in gels
with high ATP concentrations. However, the model without Brownian forces 
rendered only a qualitative description at ATP concentrations at which stable 
non-contractile {\it in vitro} active gels are prepared for rheological measurements.
Is important to emphasize that no effective temperature 
has been introduced in the active single-chain model
with Brownian forces and the motor dynamics are treated in
the same way that was done in the model that did not
include Brownian forces.

In this work it has been shown that the interplay between motor and Brownian 
forces determines the time scale for the transition from diffusive to ballistic
mass transport that has been observed in microtubule solutions 
as ATP concentration is increased \citep{sanchez2012spontaneous}. 
Moreover it has also been shown here that the breadth of relaxation spectrum 
of the active single-chain model increases
significantly when Brownian forces are taken into account. 
With respect to the predictions with the active single-chain model
without Brownian forces this broadening of the 
relaxation spectrum occurs mostly at shorter time scales, as expected.
However, for ratios between motor to Brownian forces 
of up to a hundred this broadening of the relaxation spectrum 
also has a significant influence in the shape of the dynamic 
modulus at lower frequencies, close to the longest relaxation time of the gel. 
This wider relaxation spectrum of the active single-chain 
model with Brownian forces represents better the observations
in microrheology experiments in actomyosin gels \citep{mizuno2007nonequilibrium}.  

From a practical perspective the results presented here shed light on
how the rheological response of active gels changes with varying ATP concentration. 
Experimentally, active gels with low concentrations of ATP are usually prepared to 
achieve stable gels where active steady states that do not 
phase separate can be studied. Moreover, failed active transport in actomyosin gels due to 
ATP depletion has been shown to be implicated in neurodegeneration and other diseases 
\cite{hirokawa2010molecular}. Detailed insight into how the relative importance
of Brownian forces affects the transport properties of active gels 
will eventually lead to a better understanding of these type of diseases.

\begin{acknowledgement}

This research was funded by CONICYT under FONDECYT 
grant number: 11170056. The author also thanks computational support 
from the National Laboratory for High Performance Computing (NLHPC, ECM-02).

\end{acknowledgement}

\begin{suppinfo}
%
%
The following files are available free of charge.
\begin{itemize}
  \item suppinfo.pdf: 
  Examples for the explicit forms of the transition rate matrices for $N=2$ and 
$N=3$ are given.
\end{itemize}

\end{suppinfo}


\providecommand{\latin}[1]{#1}
\makeatletter
\providecommand{\doi}
  {\begingroup\let\do\@makeother\dospecials
  \catcode`\{=1 \catcode`\}=2 \doi@aux}
\providecommand{\doi@aux}[1]{\endgroup\texttt{#1}}
\makeatother
\providecommand*\mcitethebibliography{\thebibliography}
\csname @ifundefined\endcsname{endmcitethebibliography}
  {\let\endmcitethebibliography\endthebibliography}{}

\end{document}